\documentclass[aps,prb,twocolumn,nofootinbib,floatfix,superscriptaddress]{revtex4-2}
\usepackage{physics}
\usepackage{graphicx}
\usepackage{dcolumn}
\usepackage{subfigure}
\usepackage{dsfont}
\usepackage[english]{babel}
\usepackage[utf8]{inputenc}
\usepackage[T1]{fontenc}
\usepackage{mathrsfs}
\usepackage{amsfonts}
\usepackage{bbm}
\usepackage{bm}
\usepackage{enumerate}
\usepackage[dvipsnames]{xcolor}
\usepackage{subdepth}
\usepackage{fnpct}
\usepackage{booktabs}
\usepackage{multirow}
\usepackage[math]{cellspace}
\usepackage{comment}
\usepackage[normalem]{ulem}
\usepackage{cancel}

\usepackage[colorlinks,
linkcolor=BrickRed,
citecolor=MidnightBlue,
urlcolor=MidnightBlue,
bookmarks=true,        
bookmarksopen=true,    
bookmarksnumbered=true,
]{hyperref}

\def\ket|#1>{| #1 \rangle}
\def\bra<#1|{\langle #1 |}
\def\<{\langle}
\def\>{\rangle}
\def\({\left(}
\def\){\right)}

\def\R{\mathbb{R}}

\def\nn{\nonumber}
\def\eps{\varepsilon}

\def\bS{\mathbf{S}}
\def\bcS{\bm{\mathcal{S}}}
\def\bfS{\bm{\mathfrak{S}}}
\def\fS{\mathfrak{S}}
\def\cS{\mathcal{S}}

\def\FF{\text{FF}}
\def\Heis{\text{Heis}}

\begin{document}
\title{Entanglement Halos}

\author{Nadir Samos Sáenz de Buruaga} \affiliation{CeFEMA, LaPMET,
  Instituto Superior Técnico, Universidade de Lisboa, Lisboa,
  Portugal.}

\author{Silvia N. Santalla} \affiliation{Dto. Física \&\ GISC,
  Universidad Carlos III de Madrid, Spain.}

\author{Germán Sierra} \affiliation{Instituto de Física Teórica,
  UAM-CSIC, Universidad Autónoma de Madrid, Cantoblanco, Madrid,
  Spain.}
  \affiliation{Kavli Institute for Theoretical Physics, University of California, Santa Barbara, CA 93106, USA}

\author{Javier Rodríguez-Laguna} \affiliation{Dto. Física Fundamental,
  Universidad Nacional de Educación a Distancia (UNED), Madrid,
  Spain.}

\begin{abstract}
We introduce the concept of {\em entanglement halos} ---a set of
strongly entangled distant sites within the ground state of a quantum
many-body system. Such halos emerge in star-like systems with
exponentially decaying couplings, 
as we show using both free-fermions and the spin-1/2 antiferromagnetic
Heisenberg model. Depending on the central connectivity, entanglement
halos may exhibit trivial and non trivial symmetry-protected topological features. Our findings highlight how geometry and connectivity can generate complex
entanglement structures with rich physical content, which can be experimentally accessible via state-of-the-art technologies.
\end{abstract}
\maketitle

{\em Introduction.---}
The integration of quantum information \cite{Nielsen.10} concepts into condensed matter has opened a new fertile, interdisciplinary, and rapidly evolving field, bridging diverse areas such as quantum technologies, quantum matter and optics \cite{Zeng.19}, and even more fundamental ones such as quantum gravity \cite{Harlow.16}. At the heart of many of these developments lies quantum entanglement-- the most striking manifestation of non-classical correlations \cite{Adesso.16}.  
\par
Indeed, entanglement underpins the pararellism in operations that may allow quantum speedup \cite{Arute.19}, it guides in the distinction of phases of matter beyond the Landau paradigm -- such as topological \cite{Wen.13} and symmetry-protected topological (SPT) \cite{Schuch.11,Chen.13} orders -- and through the holographic principle -- non-gravitational systems with sufficient entanglement may exhibit the characteristics signatures of quantum gravity, suggesting that entanglement can play a role in emergent spacetime geometries \cite{VanRaamsdonk.10,Brown.23}.
\par
A central principle in the study of quantum many-body systems is the area law for entanglement entropy (EE)  \cite{Amico.08}, which, in one dimension, implies that the EE remains independent of the system size.  This property underlies the efficient description of ground states using matrix product states (MPS) \cite{Cirac.21}. A notable exception to the 1D area law arises in systems whose low-energy behavior is governed by conformal field theory (CFT), where the EE exhibits a logarithmic violation of the area law \cite{Vidal.03,Calabrese.04}.  Even stronger violations can occur in certain inhomogeneous spin chains, where the entanglement entropy grows more rapidly with system size.


A paradigmatic example is the {\em rainbow chain} \cite{Vitagliano.10, Ramirez.14b, Ramirez.15} of $2L$ fermionic sites or spins whose couplings decay exponentially from the center towards the extremes. For strong inhomogeneity, the strong disorder renormalization group  \cite{Dasgupta.80, Igloi.18} show that symmetrically placed sites around the center become strongly entangled, thus yielding a set of concentric valence bonds. 

The emergence of these nonlocal Bell states constitutes the simplest example of what we call the {\em entanglement halos}. It should be noted that the concept of a \textit{halo} is well established in physics, denoting a ring- or shell-like distribution. In astrophysics, it refers to the real-space glow resulting from incoherent small-angle scattering off interstellar dust \cite{Draine.03}. In ultracold atoms physics, a halo designates the momentum-space shell formed when two dilute Bose–Einstein condensates collide and scatter coherently, producing correlated atom pairs \cite{Perrin.07,Shin.19,Shin.20}. Here we designate the entanglement halo as a well-defined entanglement structure within the ground state among certain constituents that do not need to be spatially close. 

In this letter, we explore the emergence of multiparticle entanglement halos in star-graph geometries due to the inhomogeneity of the local interactions, delving in the influence of the underlying geometry and connectivity of the system, which originates trivial and non-trivial SPT phases. We discuss that such engineered quantum systems can be realized with current state-of-the-art technologies, opening new avenues to probe entanglement structures in synthetic geometries.

{\em Ring and site stars.---} We consider a star-graph geometry
consisting of $n_B$ chains, referred to as {\em branches}, each
comprising $\ell$ nodes that host local degrees of freedom such as spins or fermions. We shall show that their physical properties depend
critically on how the branches are connected, rather than on the
nature of the degrees of freedom.

\begin{figure}[tbp]
\includegraphics[width=0.9\linewidth]{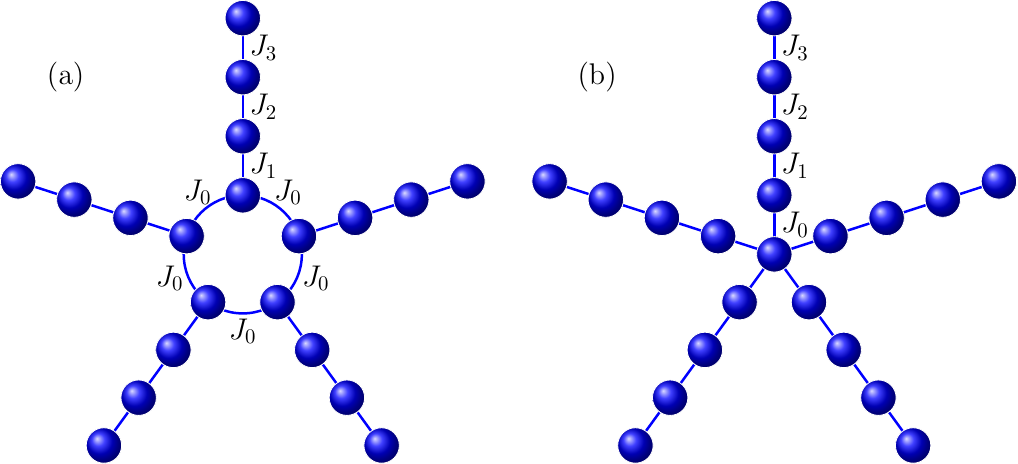}
  \caption{Star geometries of $n_B=5$ branches and $\ell=4$.(a)
    Ring-star with $N=n_B\ell=20$ and (b) Site-star comprising
    $N=n_B\ell+1=21$ sites.}
\label{fig:illust}
\end{figure}

We distinguish between two types of
graph geometries. In the {\em ring-star} geometry, the $n_B$ branches are connected
through an internal ring that links their innermost sites, as
illustrated in Fig.~\ref{fig:illust}(a). In contrast, the {\em
  site-star} geometry connects the $n_B$ branches through an additional
central site, as shown in Fig.~\ref{fig:illust}(b). Consequently, the
ring-star (site-star) configuration contains $N=n_B\ell$
($N=n_B\ell+1$) sites.

Thus, each star consists of concentric {\em rings}, which are indexed
outwards as $r \in \{0,\ldots,\ell\}$, with $r=0$ reserved for the
central site (if it exists). Branches, on the other hand, are labeled
as $p\in \{1,\ldots,n_B\}$. The $r$-th site along the $p$-th branch is
labeled as $i=(r-1)n_B+p$, and the index $i=0$ is reserved for the
central site (if it exists).

Such stars have been considered previously in the literature, both in
the homogeneous \cite{Calabrese.12,Crampe.12} and the disordered case
\cite{Juhasz.18}. Also, the possibility of a single detached ring was
considered within a particular case in Ref. \cite{Santos.18}. Star-like lattices provide a natural framework in order to extend the rainbow
chain to models with SU(3) symmetry, in which color singlets play the
role of Bell pairs \cite{Zhang.23}.

{\em Models. ---}
We analyze two representative models, interacting and noninteracting. Given a graph with $N$
sites,  we consider a  particle-conserving free-fermionic  model given by the Hamiltonian:
\begin{equation}
H_{\FF}=-\sum_{\<i,j\>} J_{ij}\, c^\dagger_i c_j\ + \ \text{h.c.},
\label{eq:ham_ff}
\end{equation}
where $\<i,j\>$ denotes the neighborhood relation in the graph, $J_{ij}\in\R^+$, and
$c^\dagger_i$ ($c_j$) denote the creation (annihilation) operators on
sites $i$ ($j$), which fulfill the anticommutation relations,
$\{c_i,c^\dagger_j\}=\delta_{ij}$.Through this work, we analyze the many-body ground state (GS) at half filling \cite{note1}.

In the interacting case, we consider the antiferromagnetic (AF) spin-1/2 Heisenberg model, defined by the Hamiltonian
\begin{equation}
  H_\Heis=\sum_{\<i,j\>} J_{ij}\, \bS_i \cdot \bS_j,
  \quad \bS_i=(\sigma^x_i,\sigma^y_i,\sigma^z_i).
\label{eq:ham_heis}
\end{equation}

{\em Inhomogeneity.---}
We consider strongly inhomogeneous couplings with radial symmetry $J_{ij}=J_r$ whenever
$i$ and $j$ belong to rings $r$ and $r+1$, as illustrated in
Fig.~\ref{fig:illust}. In addition, we enforce  a strong hierarchy on the coupling strengths  $J_0 \gg J_1 \gg J_2 \gg \cdots \gg J_{\ell-1}$. Specifically, inspired by previous work \cite{Ramirez.15,Laguna.16},
we shall consider an exponential decay of the couplings strength,
\begin{align}
  J_r&=\begin{cases}
  J_0=e^{-h/2},\\
  J_{r>0}=e^{-h r},\ r\in\{1,\cdots,\ell-1\},
  \end{cases}
  \label{eq:couplings}
\end{align}
where $h \in \mathbb{R}^+$ is an inhomogeneity parameter that reduces to the homogeneous case if $h=0$.

In the case of the ring-star geometry with $n_B =2$ legs, i.e., a chain, the ground state of Hamiltonians \eqref{eq:ham_ff}  and \eqref{eq:ham_heis} with couplings \eqref{eq:couplings} can be determined in the $h\gg1$ limit using the strong-disorder renormalization group (SDRG) approach. Originally developed for disordered chains, this RG remains highly accurate as long as the system exhibits a strong energy hierarchy. The method proceeds by iteratively decimating low-energy degrees of freedom, assumed to be spatially localized. These degrees of freedom are approximately disentangled from the rest of the system, resulting in a GS that is essentially a product state. In this case, with this method one obtains the rainbow state \cite{Vitagliano.10, Ramirez.14, Ramirez.14b, Ramirez.15}.


\begin{figure}[tbp]
\includegraphics[width=0.9\linewidth]{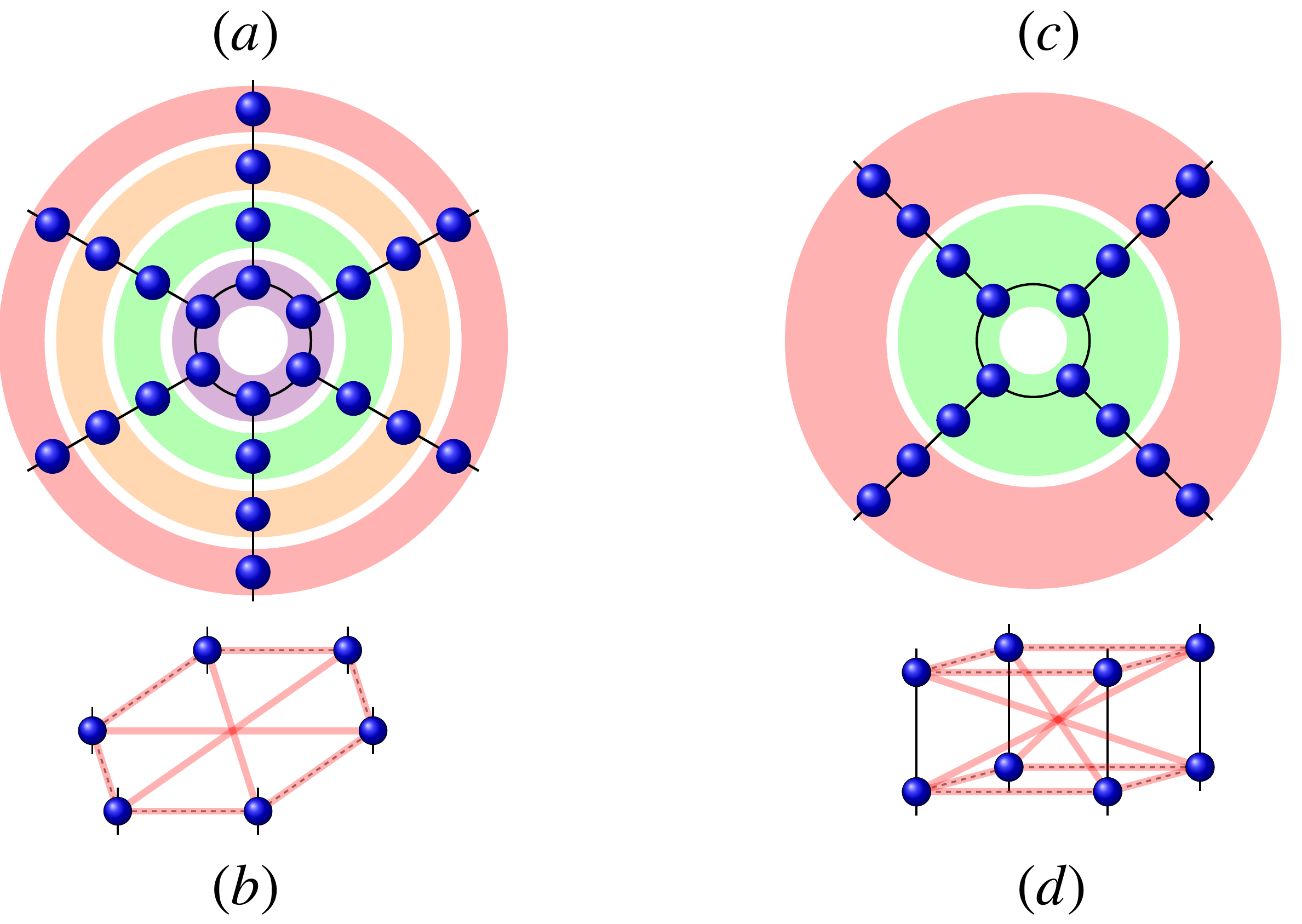}
  \caption{Entanglement structure of the free-fermionic ring-stars.
    (a) Entanglement halos in a $n_B=2$ mod 4 free-fermionic
    ring-star. Sites along each ring get entangled, while different
    rings factorize. (b) Folding the ring-star yields a tower of
    rings, whose internal entanglement pattern shows connections
    towards odd-neighbors. (c) Entanglement double halos in a $n_B=0$
    mod 4 free-fermionic ring-star. Sites along each ring pair will
    get entangled. (d) Two rings along the folded tower, showing the
    internal entanglement pattern: odd neighbors in the same floor,
    even neighbors across floors.}
  \label{fig:rcs}
\end{figure}

{\em Single-ring entanglement halos.---} We start by considering a
ring-star shown in Fig.~\ref{fig:illust}(a) and assume a strong inhomogeneity regime $h\gg 1$.
For both the interacting and non-interacting models, the central ring,
$r=1$ forms a closed chain, which is described by the Hamiltonian
\begin{equation}
\begin{split}
  H^{(1)}_\FF &= -J_0\sum_{p=1}^{n_B} c^\dagger_p c_{p+1}
  + \text{h.c.}, \quad c_{n_B+1} \equiv c_1,\\
  H^{(1)}_\Heis &= J_0\sum_{p=1}^{n_B} \bS_p \cdot \bS_{p+1},
  \quad \bS_{n_B+1} \equiv \bS_1.
  \label{eq:ham_central_ring}
  \end{split}
\end{equation}
We first consider the case in which the GS $\ket|\psi^{(1)}_0>$ of $H^{(1)}$ is unique. This is the case when $n_B$ is even for the Heisenberg model or  $n_B\not\equiv 0 \mod 4$ for the free-fermion model. The coupling hierarchy, $J_0\gg J_1$, ensures that the innermost ring described by $\ket|\psi^{(1)}_0>$  detaches from the rest of the system, weakly affecting neighboring sites in the ring $r=2$.

Following the RG spirit, we show in the SM that, for our current choice of inhomogeneities \eqref{eq:couplings}, at each RG step $r$, one can detach the inner ring $r$, described by the GS $\ket|\psi^{(r)}_0>$ of an effective Hamiltonian $H^{(r-1)}$ obtained via second-order perturbation theory. As a result, the many-body GS is approximately described by a product state.
\begin{equation}
 \ket|\Psi_{\text{GS}}>\approx \bigotimes_{r=1}^\ell\ket|\psi^{(r)}_0>,
\label{eq:product_halos}
\end{equation}

where we identify each $\ket|\psi^{(r)}_0>$ as forming an {\em entanglement halo}, as illustrated in Fig.~\ref{fig:rcs}(a).

\begin{figure}
\includegraphics[width=\linewidth]{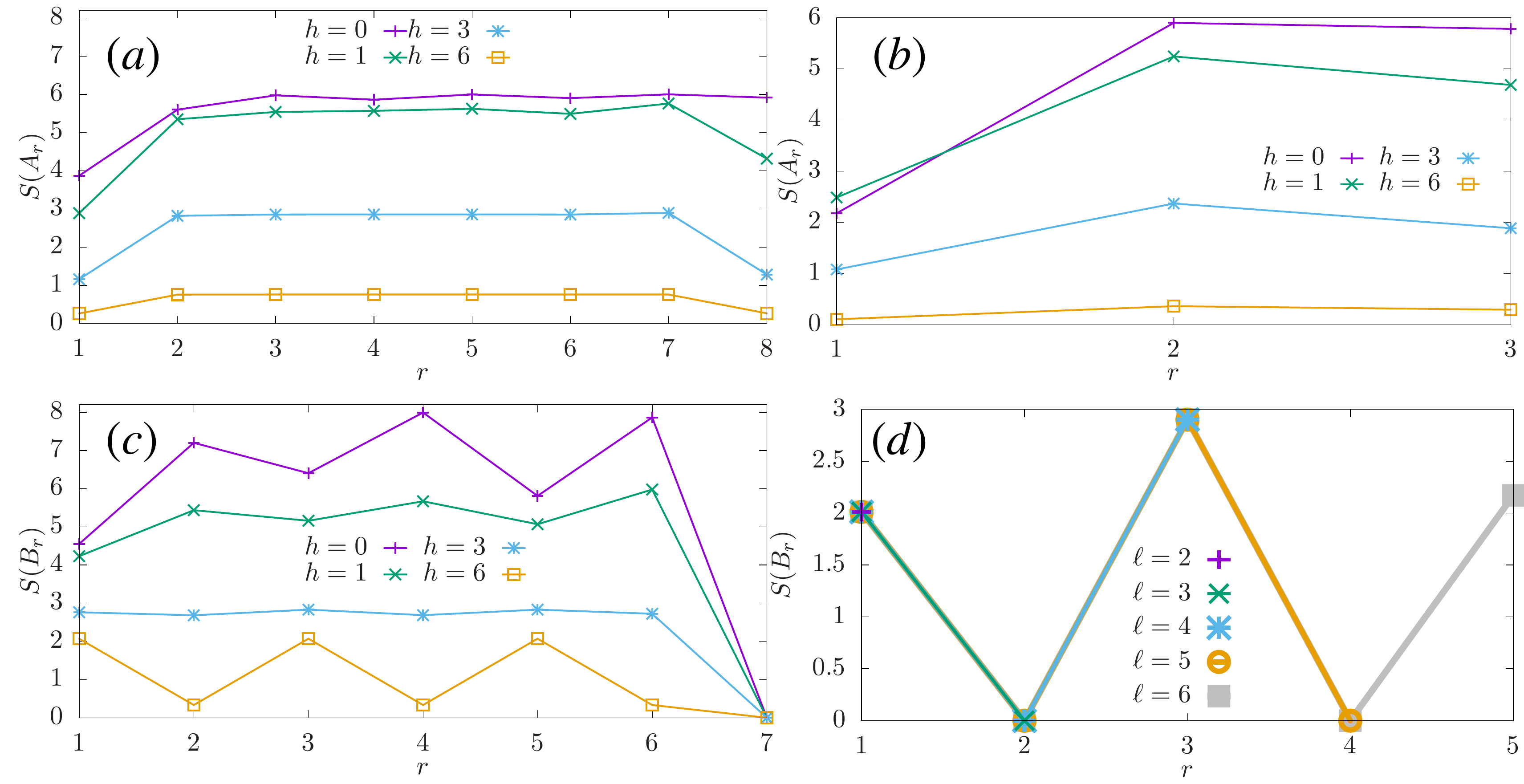}
  \caption{Top. Ring-stars presenting single-ring entanglement halos, with
    $n_B=6$. EE of the $r$-th ring, $S(A_r)$, for (a) free-fermionic
    with $\ell=8$, and (b) Heisenberg ring-stars with $\ell=3$, using
    different values of $h$. Bottom. Ring-stars presenting double-ring entanglement halos. EE of
    central blocks, containing up to the $r$-th ring, $S(B_r)$, for
    (c) the free-fermionic case with $n_B=8$, $\ell=7$, and different
    values of $h$, and (d) the Heisenberg case with $n_B=3$ and
    $\ell=6$ and $h=6$. The double-ring structure is made manifest in
    the alternation in $S(B_r)$, which vanishes for even $r$.}
  \label{fig:halos}
\end{figure}

Numerical evidence for the validity of these results is shown in Fig.
\ref{fig:halos}, where we plot the EE of the $r$-th ring $A_r$
for both (a) free-fermionic and (b) Heisenberg ring-stars using
$n_B=6$ and different values of $h$ and $\ell$. Notice that $S(A_r)$ decreases as $h$ increases approaching zero in the limit $h\gg1$, which supports the factorization in \eqref{eq:product_halos} . 
 
{\em Double-ring entanglement halos.---} Whenever the GS of $H^{(1)}$
is degenerate, e.g. for Heisenberg ring-stars with odd $n_B$ (due to
geometric frustration) or free-fermionic stars with $n_B\equiv $0 mod 4,
the entanglement structure becomes more involved. As shown in the SM,
the degree(s) of freedom associated with the degeneracy are
transferred to the $r=2$ ring, thus yielding a more complex
Hamiltonian $H^{(2)}$. Yet, the GS of $H^{(2)}$ is unique, yielding a well-defined global state $\ket|\psi^{(1,2)}>$ for the pair of rings, which detaches again from the rest of the system. The effective
Hamiltonian $H^{(3)}$ for the third ring is proportional to the first
one, $H^{(1)}$, and the double-halo structure reappears, giving rise to a product state many-body GS.
\begin{equation}
  \ket|\Psi_{\text{GS}}>\approx \bigotimes_{m=1}^{\ell/2}
  \ket|\psi_0^{(2m-1,2m)}>,
\end{equation}
Evidence for this structure is shown in Fig. \ref{fig:halos},
where we observe the EE of central blocks $S(B_r)$, which contain all
sites inside ring $r$. We observe an alternating pattern, both for (c)
free-fermionic ring-stars ($\ell=7$, $n_B=8$) or (d) Heisenberg
ring-stars ($n_B=3$, $\ell=6$, $h=6$). We should emphasize that
double-ring halos for $n_B=0$ mod 4 free-fermionic ring-stars are
associated to an accidental degeneracy, which can be lifted with an
internal magnetic flux through the central ring.

In summary, the GS of ring-stars in strong inhomogeneity regime is always a product state of either
single-ring or double-ring halos, showing a topologically trivial
structure.

{\em Twisted entanglement halos.---} We follow by considering  the site-star
geometry shown in Fig.~\ref{fig:illust}(b) and evaluate the GS of both
models with couplings given by Eq.~\eqref{eq:couplings} in the regime of $h\gg1$. The inner ---most energetic--- subsystem is
given now by the Hamiltonians
\begin{equation}
  \begin{split}
    H^{(1)}_\FF &= -J_0\sum_{i=1}^{n_B} c^\dagger_0 c_{i} + \text{h.c.},\\
    H^{(1)}_\Heis &= J_0\sum_{i=1}^{n_B} \bS_0 \cdot \bS_{i},
  \end{split}
  \label{eq:site_system}
\end{equation}
for fermions and spins, respectively. 
In similarity with the double-ring halo, the GS of Eq.~\eqref{eq:site_system} presents
degeneracy. Indeed, we can write an effective Hamiltonian for the
second ring, $H^{(2)}$, which includes the degrees of freedom
associated to this degeneracy. Yet, in opposition to the double-ring
halo case, now $H^{(2)}$ also presents degeneracy, which extends along
the RG procedure. Thus, the GS of the site-star is not factorizable.

\begin{figure}
\includegraphics[width=0.9\linewidth]{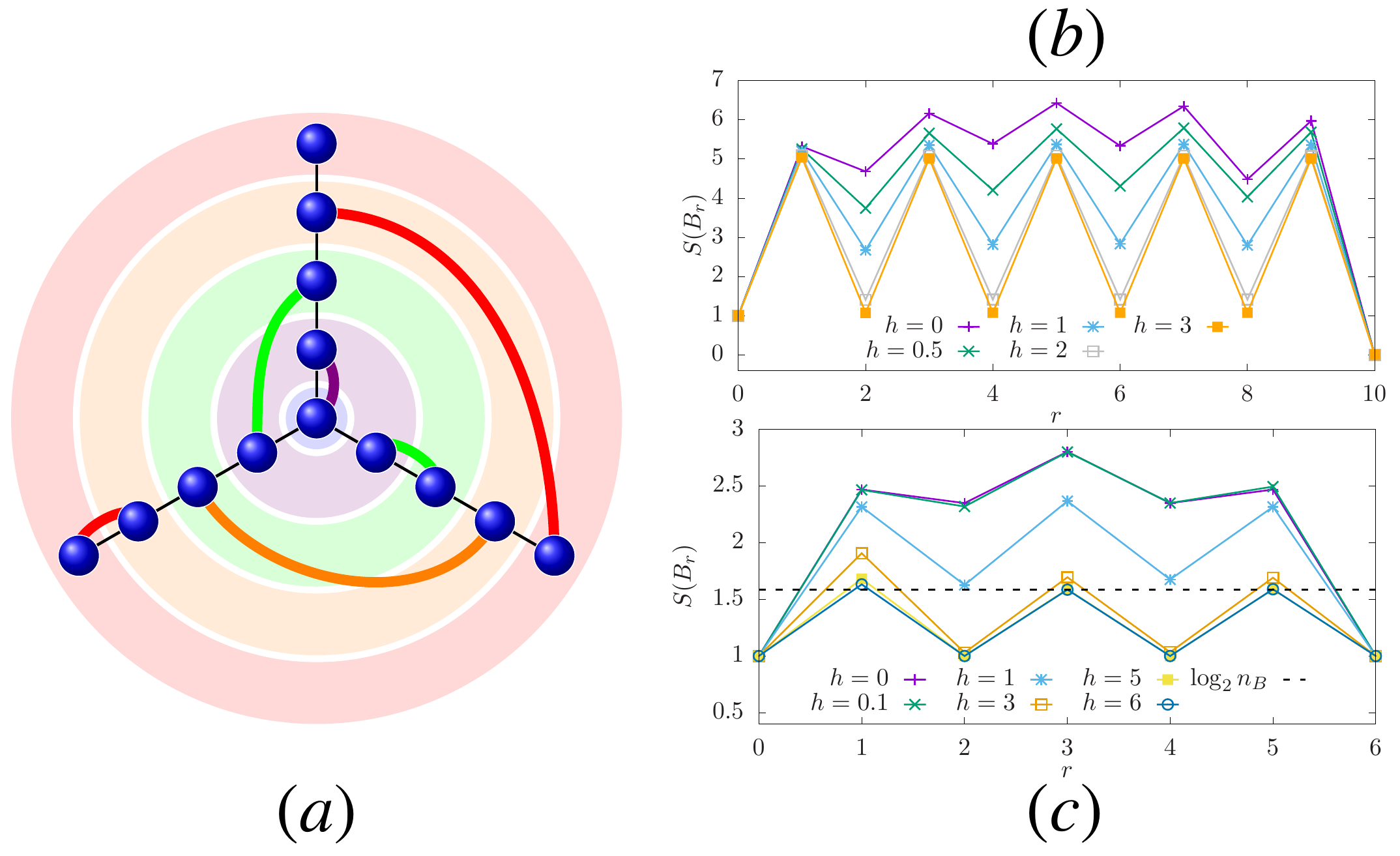}
\caption{Twisted halos in site-stars. (a) Diagram showing a typical
  bond structure. Notice that sites never establish a bond within the
  same ring. (b) EE of central blocks, $S(B_r)$, for different values
  of $h$ for a free-fermion system with $n_B=6$ and $\ell=10$ rings.
  (c) Same observable for a Heisenberg system with $n_B=3$ and
  $\ell=7$.}
\label{fig:ee_scs}
\end{figure}
 
Let us first consider the free-fermionic case. The single-body
energies of Hamiltonian $H^{(1)}$ are $\pm J_0\sqrt{n_B}$, plus
$n_B-1$ zero modes. Consequently, the many-body GS manifold presents a
large degeneracy, $2^{n_B-1}$. A single-body real-space RG (see Ref. \cite{Samos.19} ) carries the
$n_B-1$ zero modes to the $r=2$ ring and aggregates them to the
effective Hamiltonian $H^{(2)}$, which only has a single zero mode,
and then the pattern repeats itself. Therefore, the EE of odd central
blocks, $S(B_{2m-1})=n_B-1$, and the EE of even central blocks is
$S(B_{2m})=1$, as shown numerically in Fig. \ref{fig:ee_scs} (a) for a
site-star with $\ell=10$ and $n_B=6$. Since the Hamiltonian is invariant under time-reversal, particle-hole and it enjoys sublattice symmetry, it belongs to the BDI class\cite{Altland.97}, sames as the paradigmatic Su–Schrieffer–Heeger (SSH) model \cite{Su.79}.

This entanglement structure can be physically understood as follows: the central site is entangled with one site in the first ring, while the remaining $n_B - 1$ sites in ring  $r = 1$  are maximally entangled with sites in the second ring. Only one site on the ring $r = 2$ then connects to ring $r = 3$, etc. In stark contrast to the ring-star case, sites within each ring remain completely uncorrelated, while displaying strong entanglement across adjacent rings. This defines a reversed entanglement pattern, which we refer to as {\em twisted halos}: structures internally detached but maximally entangled with neighboring layers, as illustrated in Fig.~\ref{fig:ee_scs}(b).

Let us now consider the Heisenberg site-star. Exploiting the SU(2)
symmetry of the model, we write $H^{(1)}$ in terms of Casimir
operators,
\begin{equation}
  H^{(1)}_\Heis = \frac{J_0}{2}\left(\bfS_1^2-\bS_0^2-\bcS_1^2\right),
  \label{eq:casimir_heis}
\end{equation}
where $\bcS_1 \equiv \sum_{i=1}^{n_B}\bS_i$ denotes the composition of
all spins in the first ring, $r=1$, and $\bfS_1 \equiv \bS_0+\bcS_1$
is obtained by further addition of the innermost spin. In the above
form, it is clear that the minimal energy of
Eq.~\eqref{eq:casimir_heis} is obtained when $\cS_1=n_B/2$ is maximal
and $\fS_1=(n_B-1)/2$ is minimal, leading to a GS manifold of
dimension $n_B$. As detailed in the SM, we devise an RG procedure in which each step $r$ yields an effective Hamiltonian
describing the spins in ring $r+1$ coupled with an effective spin
$\bfS_r$ built from the inner spins,
$H^{(r+1)}=\tilde{J}_r\bfS_r\cdot\bcS_{r+1}$, where
\begin{equation}
  \bcS_r \equiv \sum_{i=1}^{n_B}
  \bS_{(r-1)n_B+i},\quad \bfS_r \equiv \bfS_{r-1}+ \bcS_r.
\end{equation}
Hence, each RG step requires that all spins in ring $r+1$ combine into
a maximal total spin $\cS_{r+1}$. Thus, the Heisenberg twisted halos
are not internally disentangled, because the spins-1/2 within each
ring fulfill a condition: they add up to a total spin of $n_B/2$.
However, analogously to the free fermion case, each halo is coupled to
a spin $\bfS_r$ composed of all inner halos, resembling a Matryoshka
doll. Energy minimization imposes this spin to be minimal, resulting
in an alternation $\fS_r=(n_B-1)/2$ (resp. $\fS_r=1/2$) for odd (even)
RG steps $r$. Furthermore, the effective coupling
$\tilde{J}_m=\xi^{(m)} J_m$ is fully determined by Wigner-Eckart's
theorem, yielding $\xi^{(m)}=(n_B+2)/(n_B+1)$ (resp. $(n_B+2)/(3n_B)$)
for odd (even) steps.

The nested structure of the renormalized spins $\bfS_r$ becomes more
transparent if it is written in terms of a Matrix Product State (MPS)
of $\ell$ halos with an additional left boundary spin $S_0=1/2$ (see
details in the SM)
\begin{equation}
  \ket|\fS_\ell>
  =\sum_{s_0 \cdots \cS_\ell}
  \Gamma^{s_0}\Gamma^{\cS_1}\tilde{\Gamma}^{\cS_1}\cdots
   \tilde{\Gamma}^{\cS_{\ell-1}}\Gamma^{\cS_\ell}
  \ket|s_0 \cS_1 \cS_2 \dots \cS_\ell>,
\end{equation}
where the bond dimension is given by the virtual spins $\fS_r$,
alternating therefore between 2 and $n_B$. Hence, the MPS is defined
by two types of tensors, $\Gamma^{\cS_m}$ and
$\tilde{\Gamma}^{\cS_m}$, with respective dimensions $2\times n_B$ and
$n_B\times 2$, obtained from the corresponding Clebsch-Gordan
coefficients. For all values of $n_B$ we observe that pairs of consecutive rings have spin-1, and their effective MPS corresponds to the Haldane phase. The $n_B=2$ case is special in that $\Gamma^{\cS_m}=\tilde{\Gamma}^{\cS_m}$ and single rings present the AKLT structure \cite{Affleck.87}.
Therefore, the EE between the neighboring rings alternates between $S(B_{2m})=1$
and $S(B_{2m-1})=\log n_B$, since each halo has
total spin $n_B/2$. This prediction is numerically checked in Fig.
\ref{fig:ee_scs} (c), where $S(B_r)$ is shown as a function of $r$ for
different values of $h$ for a Heisenberg site-star with $n_B=3$ and
$\ell=7$.

{\em Observables.---} EE measures are still challenging, even though
they are slowly becoming commonplace \cite{Islam.15, Lin.24}. Thus, it
is convenient to provide other experimentally accessible observables
in order to check the validity of our theoretical predictions
regarding the entanglement halos.

\begin{figure}
\includegraphics[width=0.9\linewidth]{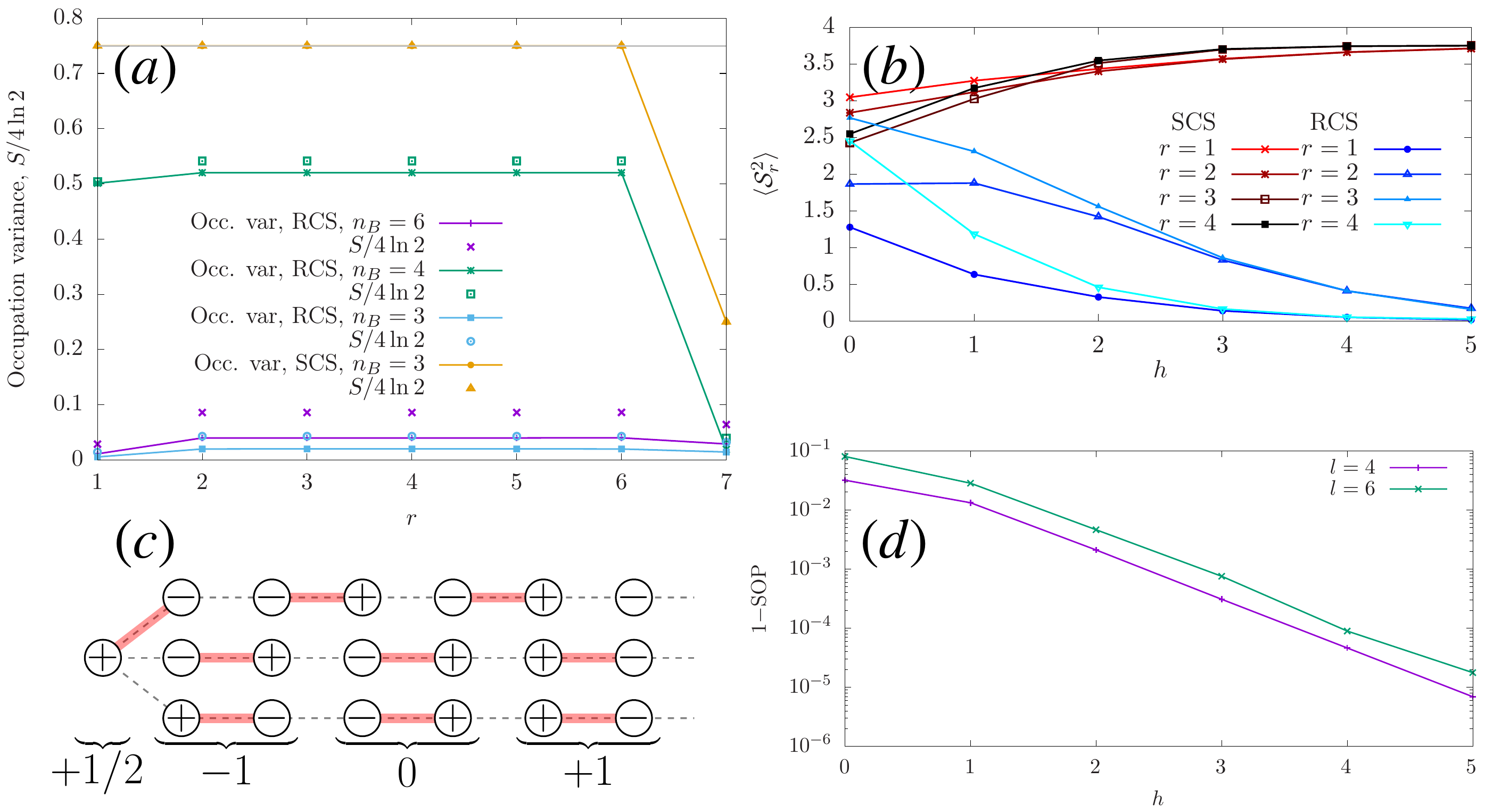}
  \caption{Observables which highlight the different phases. (a) Variance of the number of fermions on each ring, along
    with the EE divided by $4\log 2$. The horizontal dashed line at
    $3/4$ marks the theoretical prediction for site-centered rings
    with $n_B=3$. (b) Expected value of $\<S_r^2\>$ for each ring of a
    site-star (ring-star) with $n_B=3$ ($n_B=4$) as a function of $h$.  In the strong-inhomogeneity, it converges to  $S_r^2=15/4$  ($S_r^2=0$). (c) Illustration of
    the SOP and the Haldane phase within a $n_B=3$ site-star. (d)
    1-SOP for ring pairs of a site-star with $n_B=3$ and $\ell=4$ and
    $6$, also as a function of $h$, showing an exponential decay
    towards zero.}
  \label{fig:observables}
\end{figure}

For free-fermionic systems, the variance in the number of particles
inside a region is a good proxy measure to its EE \cite{Klich.06,
  Laguna.20}, based on the bound $S_A\geq 4\log 2\, \sigma_{N_A}^2$.
This variance is shown in Fig.~\ref{fig:observables} (a) for different ring
and site stars, along with the entropy of the corresponding blocks
divided by $4\log 2$, highlighting how tight the bound is in practice.

In the Heisenberg case, the total spin of certain blocks of spins highlight the different phases. As shown in Fig. \ref{fig:observables} (b), the entanglement halos of ring-stars with even $n_B$ tend to form a singlet with $S_r^2=0$, whereas in site-star geometry, each halo exhibits maximal spin
$S_r^2=(n_B/2)$. 

Yet, the hallmark of the Haldane phase is given by the string order
parameter (SOP) \cite{DenNijs.89}. The total spin of a consecutive pair of twisted halos, $\bS_{P,k}=\bcS_{2k-1}+\bcS_{2k}$, will be one,
independently of $n_B$. By measuring the $Z$-components of the
central spins and for each pair, $\{S^z_0, S^z_{P,1}, S^z_{P,2}
\cdots, S^z_{P,\ell/2}\}$, we observe that it follows a diluted
Néel pattern, i.e. a perfect sign alternation when the zeros are
removed. An illustration is provided in Fig. \ref{fig:observables}
(c), where we can see a sign assignment for the different spins-1/2 in
a $n_B=3$ site-star within a possible bond configuration and the
corresponding $S^z_{P,k}$ values. We may define the SOP in this system
as the probability of only getting diluted-Néel configurations. Fig.
\ref{fig:observables} (d) shows $1-$SOP for site-stars with $n_B=3$,
and $\ell=4$ and 6, showing an exponential decay towards zero with
$h$.

{\em Experimental implementation.---} Engineering the ring-star and
site-stars topologies along with the required decay in the couplings
poses an interesting experimental challenge. Among the available
technologies for quantum simulators, we propose the use of {\em
  synthetic dimensions} in ultracold atomic systems
\cite{Boada.12,Boada.15,Ozawa.19}, employing an internal degree of
freedom such as e.g. the hyperfine levels of a suitable atomic species
in order to represent the different nodes along a branch, with
suitably tailored Raman intensities inducing the required couplings.
Thus, each of the $n_B$ branches can be implemented inside a single
atom, which can be strategically placed in an optical lattice in order
to build the final ring-star or site-star. Another promising direction is coupling of several coplanar waveguide resonators to produce artificial photonic materials in an effective curved space \cite{Kollar.19}, including hyperbolic lattices, as required by our systems, where the inhomogeneity may be seen as a lattice proxy of spacetime curvature \cite{Laguna.17, Samos.21}.

{\em Conclusions.---} In this letter we have presented the concept of entanglement halo as the emergence of an entanglement structure among constituents which are distant according to the notion of locality given by their Hamiltonian.

We exemplified the notion using  inhomogeneous free fermion and antiferromagnetic Heisenberg spin 1-/2  arranged in two different star-graph geometries, highlighting the impact of geometry and its interplay with inhomogeneity.

The most straightforward example of entanglement halos is provided by the single and double rings appearing in ring-stars with exponentially decaying hoppings, both for free-fermionic and spins, depending on the number of branches, $n_B$ as illustrated in Fig. \ref{fig:rcs}.  These findings may have implications for quantum communication protocols based on spatially structured entanglement. However, the expected fragility of these halos under decoherence points to the need for further study regarding their robustness and controllability.

For site-star systems, the many-body GS is far from being a product state, and no subsystem is disentangled from the rest. Yet, its structure is transparent in terms of  {\em twisted halos}, in
which sites within a ring are (relatively or completely) disentangled among themselves, and strongly entangled with neighboring adjacent rings, as depicted in Fig. \ref{fig:ee_scs} (b). Crucially, twisted-halos constitute a realization of non-local SPT phases. In particular, we observe the emergence of the Haldane phase independently of the number of branches. While this phase has been previously implemented in spin-1/2 ladders \cite{White.96,Sompet.22}, our work proposes a nonlocal realization. We suggest an experimental implementation based on the use of synthetic dimensions, within the reach of state-of-the art ultracold atomic technologies, or artificial photonic lattices, along with relevant proxy observables.

Finally, although the exponential decay of the couplings may seem experimentally unrealistic, it may be possible to find optimal- polynomial- decay yielding similar phenomena. In any case, research in these many-body systems may lead to new insights into the interface of strongly-interacting systems, quantum mechanics and gravity.

\emph{Acknowledgments.---} 
SNS and JRL acknowledge the Spanish MINECO grant PID2021-123969NB-I00. NS acknowledges support from
DQUANT QuantEra II Programme supported by FCT-Portugal Grant Agreement
No. 101017733 and through the financing of the I\&D unit: UID/04540 - Centro de Física e Engenharia de Materiais Avançados. G.S. acknowledges financial support from the Spanish MINECO grant PID2021-127726NB-I00, 
the CSIC Research Platform on Quantum Technologies PTI-001, the QUANTUMENIA project 
Quantum Spain funded through the RTRP-Next Generation program under the framework of
the Digital Spain 2026 Agenda and partial support from NSF grant PHY-2309135 to the Kavli Institute
for Theoretical Physics (KITP), as well as joint sponsorship from the Fulbright Program and the Spanish
Ministry of Science, Innovation and Universities.



\clearpage
\onecolumngrid

\setcounter{figure}{0}
\renewcommand{\thefigure}{SM\arabic{figure}}%
\setcounter{page}{1}
\renewcommand{\thepage}{SM-\arabic{page}}%

\begin{center}\Large{
    \textbf{Supplemental Material for}\\
    \textbf{Entanglement halos}\vspace{2ex}}
\end{center}

\bigskip

{\em Nadir Samos Sáenz de Buruaga, Silvia N. Santalla, Germán Sierra,
Javier Rodríguez-Laguna}

\bigskip

This supplemental material provides the mathematical details
associated to the renormalization group (RG) approaches required in
order to determine the different entanglement halo structures found
along the main text. Section \ref{app:ring_star} discusses the ring-star
case, employing the free-fermionic Hamiltonian as our main example,
using both a many-body and a single-body approach.  
Section
\ref{app:site_star} introduces our RG approach for site-stars for the
Heisenberg model, proving the emergence of the Haldane phase.

\section{Ring-star geometry}
\label{app:ring_star}

We will first discuss the many-body approach for free-fermionic
systems, which can be extended to interacting systems, such as the
Heisenberg model, with the necessary changes. We expect the spin 1/2 case to have same qualitative behavior, but much more involved to treat  analytically, since it requires applying perturbation theory to the ground state of the AF Heisenberg chain. We, however, support our claims with the numerical evidence presented in the main text.

\subsection{Non-degenerate ground state}

Let us consider the free-fermionic Hamiltonian \eqref{eq:ham_ff} with
couplings given by Eq. \eqref{eq:couplings} on a ring-star with $n_B$
branches. We will require the ground state (GS) of the system to be
non-degenerate, in order to obtain single halos. This implies, in
practice, that $n_B=2$ mod 4, in absence of a central flux. If the
inhomogeneity is big enough, $h\gg1$, it is a good approximation to
focus on the subsystem composed by central ring, which contains the
strongest couplings,

\begin{equation}
  H^{(1)}=-\sum_{j=1}^{n_B}t(c^\dagger_{1,j}c_{1,j+1}+ \text{h.c.})\ ,
  \label{eq:app_h0}
\end{equation}
and treat as a perturbation their first neighbors
\begin{equation}
  H_2=-\sum_{j=1}^{n_B}J_1(c^\dagger_{1,j}c_{2,j}+ \text{h.c.}).
  \label{eq:app_h1}
\end{equation}
The GS of $H^{(1)}$ is easily obtained making use of the translational
invariance of the model and switching to momentum space
\begin{equation}
  \ket|\psi^1_{0}>=\prod_{k\in\Omega_1}d^\dagger_{1,k}\ket|0>\ , 
  \label{eq:app_gs1}
\end{equation}
with 
\begin{equation}
  d_{1,k}=\sum_{j=1}^{n_B}U_{k,j}c_{1,j}
  =\frac{1}{\sqrt{n_B}}\sum_{j=1}^{n_B}e^{ikj}c_{1,j}\ ,
  \label{eq:app_fourier}
\end{equation}
and $\Omega_1=\{m\in[-n_B/2+1,n_B/2] \ \text{ s.t }
\ \epsilon_0=-2t\cos(2m\pi/n_B)<0 \}$. We introduce the superindex $1$
to emphasize that this state belongs to the Hilbert space associated
with the level-1 operators $\{c_{1,j}\}$. Prior to studying the effect
of $H_2$, we extend the eigenstates of $H^{(1)}$: $\ket|\psi^1_k>\to
\ket|\psi^1_k, I>$, where $\ket|I>$ corresponds to a generic state
belonging to the Hilbert space spanned by the level-2 operators
$c_{2,j} \ j=1\dots n_B$.

It is easy to check that $\mel{\psi^1_0,I}{H_2}{\psi^1_0,I}=0$. The
matrix element of the degenerate second-order contribution is:
\begin{eqnarray}
B_{I,J}=\sum_{i\neq0}\sum_{I'}\frac{1}{E_0-E_i}
  \mel{\psi^1_0,I}{H_2}{\psi^1_i,I'}\mel{\psi^1_i,I'}{H_2}{\psi^1_0,J}
\end{eqnarray}
Plugging Eq.\eqref{eq:app_h1} above and expanding yields

\begin{eqnarray}
  B_{I,J}=\sum_{i\neq0}\frac{J_1^2}{E_0-E_i}\sum_{j,j'=1}^{n_B}
  \mel{I}{c_{2,j}c_{2,j'}}{J}\mel{\psi^1_0}{c^\dagger_{1,j}}{\psi^1_i}
  \mel{\psi^1_i}{c^\dagger_{1,j'}}{\psi^1_0}
  +\mel{I}{c^\dagger_{2,j}c_{2,j'}}{J}\mel{\psi^1_0}{c_{1,j}}{\psi^1_i}
  \mel{\psi^1_i}{c^\dagger_{1,j'}}{\psi^1_0} \nonumber \\
  \mel{I}{c_{2,j}c^\dagger_{2,j'}}{J}
  \mel{\psi^1_0}{c^\dagger_{1,j}}{\psi^1_i}
  \mel{\psi^1_i}{c_{1,j'}}{\psi^1_0}
  +\mel{I}{c^\dagger_{2,j}c^\dagger_{2,j'}}{J}
  \mel{\psi^1_0}{c_{1,j}}{\psi^1_i}\mel{\psi^1_i}{c_{1,j'}}{\psi^1_0}.
\end{eqnarray}
Inverting the relation Eq. \eqref{eq:app_fourier} and substituting
above we obtain
\begin{eqnarray}
  B_{I,J}=J_1^2\sum_{i\neq0}\frac{1}{E_0-E_i}
  \sum_{j,j'=1}^{n_B}\sum_{k,k'}
  \mel{I}{c_{2,j}c_{2,j'}}{J}
  \mel{\psi^1_0}{d^\dagger_{0,k}}{\psi^1_i}
  \mel{\psi^1_i}{d^\dagger_{0,k'}}{\psi^1_0}U_{k,j}U_{k',j'}\nonumber\\
  +\mel{I}{c^\dagger_{2,j}c_{2,j'}}{J}\mel{\psi^1_0}{d_{0,k}}{\psi^1_i}
  \mel{\psi^1_i}{d^\dagger_{0,k'}}{\psi^1_0}U_{k,j}U^*_{k',j'}\nonumber \\
  +\mel{I}{c_{2,j}c^\dagger_{2,j'}}{J}
  \mel{\psi^1_0}{d^\dagger_{0,k}}{\psi^1_i}
  \mel{\psi^1_i}{d_{0,k'}}{\psi^1_0}U^*_{k,j}U_{k',j'}\nonumber\\
  +\mel{I}{c^\dagger_{2,j}c^\dagger_{2,j'}}{J}
  \mel{\psi^1_0}{d_{0,k}}{\psi^1_i}
  \mel{\psi^1_i}{d_{0,k'}}{\psi^1_0}U^*_{k,j}U^*_{k',j'}.
  \label{eq:app_generic_ptheory}
\end{eqnarray}
Note that the above expression can be non-zero for those excited
states that differ in just one fermion with energy
$E_i=E_0\pm\epsilon_k$. With this consideration, we obtain the
following:
\begin{eqnarray}
  B_{I,J}=-J_1^2\sum_{j,j'=1}^{n_B}\sum_{k,k'}\frac{1}{\epsilon_k}
  \(\mel{I}{c^\dagger_{2,j}c_{2,j'}}{J}U_{k,j}U^*_{k',j'}
  -\mel{I}{c_{2,j}c^\dagger_{2,j'}}{J}{\psi^1_0}U^*_{k,j}U_{k',j'}\)\delta(k-k')
\label{eq:app_generic_ptheory_result}
\end{eqnarray}
Finally, we particularize this result for the closed chain
$U_{k,j}=e^{ijk}/\sqrt{n_B}$  and $\epsilon_k=-2t\cos(k)$:
\begin{equation}
  B_{I,J}= \frac{J^2}{2 t n_B}\frac{1}{2}
  \sum_{j,j'=1}^{n_B}\sum_{k}\(\frac{e^{ik(j-j')}}{\cos(k)}
\mel{I}{c^\dagger_{2,j}c_{2,j'}}{J}+ \ \text{h.c.}\) + \text{const.}\ ,
\end{equation}
where the additional $1/2$ compensates the double counting in the sum.
It can be shown that
\begin{equation}
  \sum_{m=-2n}^{2n+1}
  \frac{e^{i\frac{m\pi}{2n+1}\Delta}}{\cos\(\frac{m \pi}{2n+1}\)}=
  \begin{cases}
    0 \ \text{if} \ \Delta=2p, \\
    (-1)^p n_B  \ \text{if} \ \Delta=2p+1.
  \end{cases}
\end{equation}
Thus, we have obtained an effective Hamiltonian that describes a
system of free fermions lying exclusively at level-2 with alternating
long-range couplings between odd neighbors:
\begin{equation}
  H^{(2)}=J^{(2)}\(\sum_{j=1}^{n_B}
  \sum_{l=1}^{n}(-1)^{l+1} c^\dagger_{2,j}c_{2,j+2l-1}+
  \frac{(-1)^{n}}{2} c^\dagger_{2,j}c_{2,j+2n+1}+ \text{ h.c. }\),
  \quad  J^{(2)}=\frac{J_1^2}{2 t}, \quad n_B+j=n_B.
\end{equation}
Thus, the level-1 is described by the state $\ket|\psi^1_0>$
Eq.\eqref{eq:app_gs1}, and it is disentangled from the rest, yielding
a star with $N-1$ levels of $n_B$ fermions each \footnote{The factor 2
in the denominator could be surprising since it appears in the SDRG of
a Heisenberg spin $1/2$ chain and not in the $XX$ chain or a free
fermion chain by means of a Jordan-Wigner transformation. Indeed,
notice that we recover $J^{(2)}=J_1^2/t$ by setting $n_B=2$ and taking
into account that the Hamiltonian Eq. \eqref{eq:app_h0} does not
describe in this case a closed system.}.

Observe that given the strength of the couplings, Eq.
\eqref{eq:couplings}, the effective Hamiltonian obtained above
presents the strongest hopping interactions
$|J^{(2)}|=e^{-\frac{3h}{2}}/2$. Hence, in the next RG step, we focus
on this system and its neighbor level-3,
\begin{equation}
    H_3=-\sum_{j=1}^{n_B}J_1(c^\dagger_{2,j}c_{3,j}+ \text{h.c.}). 
    \label{eq:app_h2}
\end{equation}
First, we diagonalize $H^{(2)}$ by switching to momentum space, Eq.
\eqref{eq:app_fourier}, yielding
\begin{equation}
  H^{(2)}=\sum_k\epsilon(k)d^\dagger_{2,k}d_{2,k},
  \ \quad \
  \epsilon(k)=J^{(2)}\sum_{l=1}^{n}2\cos\(k(2l-1)\)+(-1)^n\cos(k(2n+1)).
\end{equation}
Imposing the boundary conditions $e^{ik(m+N)}=e^{ikm}$, it can be
shown that the single body energies are
\begin{equation}
  \epsilon(k)= J^{(2)}(-1)^n
  \cos(2 k n+k)+2\sec(k)\sin^2\(k n-\frac{\pi n}{2}\)
  \ , \quad
  k=\frac{m\pi}{2n+1},\  m=-2n+1,\dots ,2n+2.
\end{equation}
As mentioned previously, the GS is obtained filling up the negative
energies,
\begin{equation}
  \ket|\psi^2_{0}>=\prod_{k\in\Omega_2}d^\dagger_{2,k}\ket|0>,
  \quad \text{ with } \Omega_2=\{k \ \text{s.t.}\ \epsilon(k)<0 \}.
\label{eq:app_gs2}
\end{equation}
Notice that $H^{(2)}$ presents sub-lattice symmetry, meaning that
$\ket|\psi^2_{0}>$ describes a state of $n_B/2=2n+1$ fermions. Like
before, we work on $H_3$, Eq. \eqref{eq:app_h2}, using second-order
perturbation theory, Eq. \eqref{eq:app_generic_ptheory_result}. Taking
into account that
\begin{equation}
    \sum_{m=-2n}^{2n+1}\frac{e^{i\frac{m \pi}{2n+1}\Delta}}{(-1)^n
      \cos (2kn+k)+2 \sec(k)\sin^2\(kn-\frac{\pi n}{2}\)}
    =\frac{n_B}{2}\(\delta(\Delta\pm1)+\delta(\Delta\pm(n_B-1))\)\ ,
\end{equation}
we obtain the effective Hamiltonian
\begin{equation}
    H^{(3)}=J^{(3)}\sum_{j=1}^{n_B}t(c^\dagger_{3,j}c_{3,j+1}+
    \text{h.c.}), \quad J^{(3)}=-\frac{J^2_3}{J^{(2)}}\ ,
    \label{eq:app_h3}
\end{equation}
that describes a closed chain of first-neighbor couplings lying
exclusively at level-3. The level-2 is given by the state
$\ket|\psi_0^2>$, Eq. \eqref{eq:app_gs2}, which becomes detached from
the star. Observe that because of the strength couplings Eq.
\eqref{eq:app_strength_couplings} of the system, the whole process of
the RG is an iteration of the two steps detailed above.

\subsection{Degenerate ground state}

The degenerate case is more involved. In this section we will cover
the $n_B=4$ case for free-fermionic ring stars, although the procedure
can be generalized to odd number of branches and other cases.

Let us generalize the SDRG procedure in the following way. Consider a
hopping single-body system composed of two types of sites: $L_0$ (with
$n_0$ sites) and $L_1$ (with $n_1$ sites). Sites in $L_0$ are strongly
linked among themselves, with links of order $J_0$, while sites in
$L_1$ only possess links to sites in $L_0$, of order $J_1\ll J_0$. Let
$P_0$ be the projector on sites of $L_0$, and $P_1$ the projector on
sites of $L_1$. Our single-body Hamiltonian will be $H=H_0 + V$, where
$H_0$ contains the internal links of $L_0$ and $V=H_{0,1}$ contains
the links joining $L_0$ and $L_1$.

We may obtain the eigenstates of the hopping system as a perturbative
series in $V$. To zero-order, we just consider the eigenstates of
$H_0$. Due to sublattice symmetry, the negative eigenvalues will be
occupied. Let us assume, for the time being, that the only zero-modes
of $H_0$ are given by the states located on the sites in $L_1$. In
that case, which we will call {\em non-degenerate}, we can write the
effective Hamiltonian for $L_1$ as a perturbation series. To first
order, we have $\tilde H^{(1)}_1 \approx P_1 V P_1$, which is zero
because $V$ does not contain links joining sites in $L_1$. Thus, we
must proceed to second order,
\begin{equation}
\tilde H^{(2)}_1 \approx P_1 V P_0 H_0^{-1} P_0 V P_1.
\end{equation}

\medskip

Yet, in this section we will focus on the degenerate case, in which
$H_0$ has more zero modes, i.e., $\dim(\text{Ker}(H_0))>n_1$. These
modes must be considered along with the sites in $L_1$ in order to
write an effective Hamiltonian. Let $P_Z$ be the projector on all the
zero-modes of $H_0$. Thus, we consider the first-order effective
Hamiltonian,
\begin{equation}
\tilde H^{(1)}_1 \approx P_Z V P_Z,
\end{equation}
which must be diagonalized. The negative-energy modes of $\tilde
H^{(1)}_1$ must be occupied in the GS, and the positive-energy modes
must be empty. Then, what happens with the zero-modes? Let $Q_Z$
denote the projector on these remaining zero modes. We must build a
second-order perturbation theory effective Hamiltonian,
\begin{equation}
\tilde H^{(2)}_1 \approx Q_Z V (1-Q_Z) H_0^{-1} (1-Q_Z) V Q_Z.
\label{eq:H2QZ}
\end{equation}
Once we have obtained an effective Hamiltonian for $L_1$, we may
iterate the procedure. Yet, there is an important {\em caveat}: in
some cases, zero modes may travel along the renormalization procedure,
and cross the full system.

\bigskip

In order to make the calculations concrete, let $\ket|u_k>=\sum_i
U_{ki} \ket|i>$ be the $k$-th eigenvector of $H_0$, with energy
$\eps_k$. In the non-degenerate case, we have
\begin{equation}
\<i|\tilde H_1|j\> = \sum_k {\<i|V|u_k\>\<u_k|V|j\>\over \eps_k}
=\sum_{k,l,m} {U_{kl} \bar U_{km} V_{il} V_{jm}\over \eps_k},
\label{eq:Heff}
\end{equation}
where $l$ and $m$ range over sites of $L_0$. The GS on $L_0$ is built
from the negative-energy modes, and the process is then iterated. The
final GS is approximately a tensor product of GS on the different
levels,
\begin{equation}
\ket|\Psi> \approx \ket|\psi_0> \otimes \ket|\psi_1> \otimes \cdots
\ket|\psi_n>.
\end{equation}
In the degenerate case the previous expression need not be correct,
since zero-modes may hybridize between different levels. In this case,
we proceed as follows:

\begin{itemize}
\item We combine the zero-modes of $L_0$ with the sites of $L_1$, and
  write the first-order perturbation matrix for them, $\tilde
  H_1^{(1)}$.
\item We diagonalize $\tilde H_1^{(1)}$, and separate the remaining
  zero modes. 
\item The non-zero modes acquire an energy and are detached from the
  subspace. They are occupied if their energy is negative.
\item We obtain the effective second-order Hamiltonian for the
  zero-modes remaining, $\tilde H^{(2)}_1$, which may combine sites
  from $L_0$ and $L_1$.
\item We fill up the negative energy modes.
\item If any zero-mode survives, it must be upgraded to the next
  level.
\end{itemize}

Rigorously, the surviving zero-modes might disappear at higher orders
in perturbation theory. We should upgrade them only if we are sure
that they are exact, and that is only true if they are protected by a
symmetry.

The aforementioned procedure suggests its use on {\em hierarchical
  graphs} with the following conditions. The lattice sites must be
decomposed into {\em levels}, ${\cal G} = L_0 \cup L_1 \cup L_2 \cdots
\cup L_n$, with $L_k\cap L_l=\emptyset$, and $|L_k|=n_k$. Sites within
level $L_k$ with $k>0$ are not linked among themselves, i.e.
$J_{i,j}=0$ if $i, j\in L_k$ and $k>0$. Furthermore, the energies
associated to the Hamiltonian $H_{k,k+1}$ are order $O(J_k)$, and
$J_0\gg J_1 \gg J_2 \gg \cdots$. The ring-centered and site-centered
rainbow stars defined above fulfill these conditions.

\bigskip

For example let us consider the free-fermionic case with $n_B=4$. The
spectrum of the single-body Hamiltonian is degenerate, $H_0$ has two
zero modes, given as follows:
\begin{align}
  \<i|u_1\>&=\{1/2,1/2,1/2,1/2\}, \qquad \varepsilon_1 = - J_0,\nn\\
  \<i|u_2\>&=\{0,1/\sqrt{2},0,1/\sqrt{2}\}, \qquad \varepsilon_2=0,\nn\\
  \<i|u_3\>&=\{1/\sqrt{2},0,1/\sqrt{2},0\}, \qquad \varepsilon_3=0,\nn\\
  \<i|u_1\>&=\{1/2,-1/2,1/2,-1/2\}, \qquad \varepsilon_1 = + J_0.
\end{align}
Thus, modes $\ket|u_2>$ and $\ket|u_3>$ are {\em upgraded} to level 1,
and we write down the perturbation matrix $V$ for the six states.
Therefore, it contains non-zero elements. We diagonalize it, by
realizing that it splits into two $3\times 3$ blocks. Four states
acquire a first-order correction, hybridizing sites from levels 0 and 1:
\begin{align}
  \ket|V_1>={1\over\sqrt{2}}\ket|u_2> + {1\over 2}\ket|1_1> + {1\over
    2}\ket|3_1>, \qquad E_1=-J_1,\nn\\
  \ket|V_2>={1\over\sqrt{2}}\ket|u_2> + {1\over 2}\ket|1_1> + {1\over
    2}\ket|3_1>, \qquad E_2=+J_1,\nn\\
  \ket|V_3>={1\over\sqrt{2}}\ket|u_3> + {1\over 2}\ket|2_1> + {1\over
    2}\ket|4_1>, \qquad E_1=-J_1,\nn\\
  \ket|V_4>={1\over\sqrt{2}}\ket|u_3> + {1\over 2}\ket|2_1> + {1\over
    2}\ket|4_1>, \qquad E_2=+J_1.
\end{align}
Furthermore, there are two surviving zero-modes, $\ket|Z_1>$ and
$\ket|Z_2>$,
\begin{align}
  \ket|Z_1> &={1\over\sqrt{2}}\( \ket|1_1>-\ket|3_1>\)  ,\nn\\
  \ket|Z_2> &={1\over\sqrt{2}}\( \ket|2_1>-\ket|4_1>\)  .
\end{align}
We may then obtain the effective second-order effective Hamiltonian
for these two zero-modes, summing over all other states, using Eq.
\eqref{eq:H2QZ} and obtain a single non-zero matrix element,
\begin{equation}
\<Z_1|\tilde H^{(2)}_1|Z_2\>={J_1^2\over 2J_0}.
\end{equation}
Thus, we prove that $\tilde H^{(2)}_1$ is non-degenerate, and thus the
second ring {\em closes up}. The first two rings, $L_0$ and $L_1$, can
be characterized by a single wavefunction, disentangled from the rest
of the chain. 


\section{Site-star geometry}
\label{app:site_star}

Since in  the previous section we considered the free fermion model, we shall analyze here the spin 1/2 system. It is worth noting that a similar single-body RG  to the one described in the previous section can be found for the free fermion site-star.

Let us consider now a fully SU(2)-symmetric Heisenberg model on
site-stars with $n_B$ branches, using the couplings Eq.
\eqref{eq:couplings}. The most energetic subsystem is described by the
Hamiltonian
\begin{equation}
        H^{(1)} = J_0\sum_{i=1}^{n_B} \bS_0 \cdot \bS_{i},
    \label{eq:app_site_system}
\end{equation}
We denote with caligraphic $\bcS_1$ the spin obtained by
coupling all spins belonging to the level $1$
\begin{equation}
\bcS_1\equiv\sum_{i=1}^{n_B}\bS_i,
\end{equation}
and the spin $\bfS_1$ obtained by coupling the above with
the innermost spin $\bS_0$
\begin{equation}
    \bfS_1\equiv\bS_0+ \bcS_1,
\end{equation}
Equiped with them we can rewrite Eq.\eqref{eq:app_site_system} Indeed
we can rewrite Eq.\eqref{eq:app_site_system} in terms of Casimir
operators
\begin{equation}
H^{(1)}=J_0\bS_0\cdot\(\sum_{i=1}^{n_B}\bS_i\)=J_0\bS_0\cdot\bcS_1
=\frac{J_0}{2}\(\bm{\mathfrak{S}_1}^2-\bS_0^2-\bm{\mathcal{S}_1}^2\),
\label{eq:app_Heisenberg_casimir}
\end{equation}
from which it follows that the ground state energy is obtained when
$\mathcal{S}_1=n_B/2$ is maximal and $\mathfrak{S}_1=(n_B-1)/2$ is
minimal. The leading perturbation of Eq.
\ref{eq:app_Heisenberg_casimir} is
\begin{equation}
W_1=J_1\sum_{i=1}^{n_B}\bS_{i}\cdot\bS_{i+n_B},
\label{eq:app_perturbed}
\end{equation}
that must be written in terms of the effective spin
$\bfS_1$. Working in the z-component basis we can write
\begin{equation} 
  \mel{\mathfrak{S}_1}{\bS^a_i}{\tilde{\mathfrak{S}}_1} =
  \xi^a_{1,i}\mel{\mathfrak{S}_1}{\bfS^a_1}{\tilde{\mathfrak{S}}_1},
\label{eq:app_spin_cakeportion}
\end{equation}
with $a=x,y,z.$. Since we consider no anisotropy, the coefficient
cannot depend on the spin component: $\xi_{1,i}^a=\xi_{i}$. Moreover,
the site-star geometry is invariant under operations of the symmetric
group of $n_B$ elements. Hence, the coefficient cannot depend on the
branch index $\xi_{1,i}\equiv\xi_{1}$. Using the above, we can compute
$\xi^z_{1,0}$ and then use the fact that
\begin{equation}
  \xi_{1,0}+\sum_{i=1}^{n_B}\xi_{1,i}=1
  \Rightarrow \xi_1=\frac{1-\xi_{1,0}}{n_B}
\end{equation}
Hence we need to compute:
\begin{equation} 
  \xi^z_{1,0}=\frac{\mel{\mathfrak{S}_1}{\bS^z_0}{\tilde{\mathfrak{S}}_1}}
     {\mel{\mathfrak{S}_1}{\bfS^z_1}{\tilde{\mathfrak{S}}_1}}
     = \frac{1}{\mathfrak{S}_1}
     \mel{\mathfrak{S}_1}{\bS^z_0}{\mathfrak{S}}_1
\label{eq:app_z_component_cake}
\end{equation}
We can infer from the Wigner-Eckhart theorem that, since $\bm{S}$ is a
vector operator, and
\begin{equation}
  \mel{\bfS_1,\mathfrak{S}_1}{\bm{S}^z_0}{\bfS_1,\mathfrak{S}_1}
  =\braket{\bfS_1,\mathfrak{S}_1;1,0}{\bfS_1,\mathfrak{S}_1}
  \bigl\langle \bfS_1 \bigl\| \bm{S}_0 \bigr\| \bfS_1 \bigr\rangle,
\end{equation}
that $\ev{\bm{S}_0^x}=\ev{\bm{S}_0^y}=0$, meaning that
$\bm{S}_0=\alpha\bfS_0$. Then it follows that
\begin{equation}
  \alpha=\frac{\ev{\bfS_1\cdot \bm{S}_0}}{\ev{\bfS_1^2}}
  =\frac{\ev{\bm{S}_0^2}+\ev{\bcS_1}\cdot\bm{S}_0}{\ev{\bfS_1^2}}
  =\frac{\ev{\bfS_1^2}+\ev{\bm{S}_0^2}
    -\ev{\bm{\mathcal{S}_1}^2}}{2\ev{\bfS_1^2}}=-\frac{1}{n_B+1}.
\end{equation}
Now we can use the fact that
$\ev{\bm{S}^z_0}=\alpha\ev{\bfS^z_1}=\mathfrak{S}_1\alpha$ and write
Eq. \eqref{eq:app_z_component_cake} $\xi_0=-1/(n_B+1)$ and therefore
\begin{equation}
  \xi_{1,i}\equiv\xi_1=\frac{1}{n_B}\frac{n_B+2}{n_B+1}
  \label{eq:app_xi_odd}
\end{equation}
Hence, we can write the perturbation Eq. \eqref{eq:app_perturbed} as:
\begin{equation}
W_1=J_1\sum_{i=1}^{n_B}\bS_{i}\cdot\bS_{i+n_B}
= J_1\xi_1 \bfS_1\cdot
\(\sum_{i=1}^{n_B}\bS_{i+n_B}\) 
= \tilde{J}_1\bfS_1\cdot\mathbf{\bcS}_2\equiv H^{(2)},
\label{eq:app_h2_heisenberg}
\end{equation}
recovering the functional form of the unperturbed Hamiltonian Eq.
\eqref{eq:app_site_system} with a renormalize coupling
$\tilde{J}_1=J_1\xi_1$. When $\tilde{J}_1\gg J_2$, the same reasoning
that we presented above applies, and the ground state of Eq.
\eqref{eq:app_h2_heisenberg} is obtained by minimizing the total spin
$\bfS_2=\bfS_1+\bcS_2$, which yields $\mathfrak{S}_2=1/2$. The leading
perturbation is then
\begin{equation}
H^{(3)}=\tilde{J}_2\bfS_2\cdot\bcS_3.
\end{equation}
Using the same reasoning above 
\begin{equation} 
  \mel{\mathfrak{S}_2}{\bfS_1^z}{\tilde{\mathfrak{S}}_2}
  = \xi^z_{2,i}\mel{\mathfrak{S}_2}{\bfS^z_2}{\tilde{\mathfrak{S}}_2},
\label{eq:app_spin_cakeportion2}
\end{equation}
and using an analogous reasoning
$\bm{\mathfrak{S}_1}=\alpha\bm{\mathfrak{S}_2}$ with

\begin{equation}
  \alpha=\frac{\ev{\bfS_2^2}+\ev{\bfS_1^2}
    -\ev{\bm{\mathcal{S}_2}^2}}{2\ev{\bfS_2^2}}
  =\frac{1}{2}\frac{3+n^2_B-1-n_B^2-2n_B}{3}=-\frac{n_B-1}{3}.
\end{equation}
with a renormalized coupling $\tilde{J}_2=\xi_2J_2$. Using Eq.
\eqref{eq:app_spin_cakeportion},
\begin{equation}
  \xi_{2,i}\equiv\xi_2=\frac{2+n_B}{3n_B}.
  \label{eq:app_xi_even}
\end{equation}

\begin{figure}
  \centering
  \includegraphics[width=10cm]{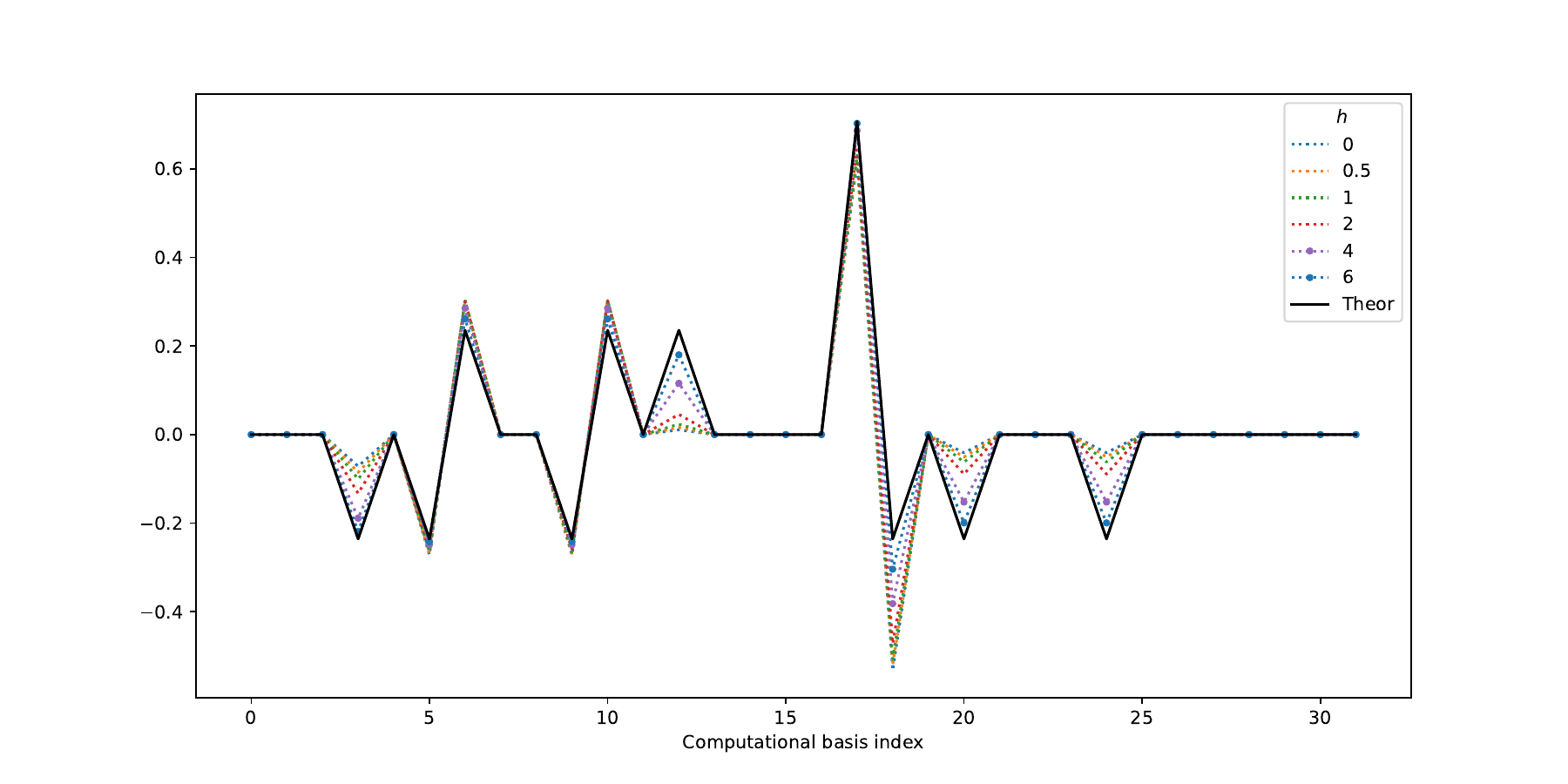}
  \caption{The black line correspond to the state Eq.
    \eqref{eq:app_gs_onehalf}. As the inhomogeneity increases, the GS
    approaches this state.}
  \label{fig:app_GS_heisenberg}
\end{figure}

To gain some intuition, consider the case of $n_B=3$ and $\ell=1$.
Following the prescription, we impose that the central spins
$\bcS_1=\bS_1+\bS_2+\bS_3$ combine, producing the maximal spin
$\mathcal{S}=\frac{3}{2}$. Then we combine $\bS_0+\bcS_1$ to yield the
minimal value, $\mathfrak{S}_1=1$.
\begin{align}
  \ket|1,1>&=\sqrt{\frac{3}{4}}\ket|-+++>
  -\sqrt{\frac{1}{12}}\(\ket|+++->+\ket|++-+>+\ket|+-++>\)\\
  \ket|1,0>&= \sqrt{\frac{1}{6}}\(\ket|-++->+\ket|-+-+>
  +\ket|--++>-\ket|++-->-\ket|+-+->-\ket|+--+>\)\\ 
  \ket|1,-1>&=\sqrt{\frac{1}{12}}\(\ket|-+-->+\ket|--+->+\ket|---+>\)
  -\sqrt{\frac{3}{4}}\ket|+--->,
\label{eq:app_gs_one}
\end{align}
where we take the convention $\ket|\bS_0,\bS_1,\bS_2,\bS_3>$,
$\bS_i=\pm\frac{1}{2}\equiv \pm$.  Using the above, we can compute
$\xi_1$
\begin{align}
  \xi_{1,0}&=\frac{\<\tilde{m}|\bS^a_0|\tilde{n}\>}
     {\<\tilde{m}|\mathfrak{S}^a_1|\tilde{n}\>}
     =-\frac{1}{4},\quad a=x,y,z ,\quad  \tilde{m},\tilde{n}=1,0,-1.\\
     \xi_{i}&=\frac{\<\tilde{m}|\bS^a_i|\tilde{n}\>}
        {\<\tilde{m}|\mathfrak{S}^a_1|\tilde{n}\>}
        =\frac{5}{12},\quad \quad i=1,2,3.
\end{align}
Observe that $\sum_{i=0}^{3}\xi_{i,1}^a=1$ $a=x,y,z$ as a consequence
of preserving $SU(2)$ and it is the value given by Eq.
\eqref{eq:app_xi_odd}. Also, let us add an additional site yielding a
total $\bfS_1+\bS_4$ spin $1/2$. In Fig. \ref{fig:app_GS_heisenberg}
we plot the ground state in the computational basis for different
values of the inhomogeneity parameter $h$ and the RG prediction
\begin{align}
  \ket|1/2>=\sqrt{\frac{1}{2}}\ket|-+++->&
  +\sqrt{\frac{1}{18}}\(-\ket|+++-->-\ket|++-+->-\ket|+-++->\right.\nn\\
  &-\ket|-++-+>-\ket|-+-++>-\ket|--+++>+\ket|++--+>\nn\\
  &\left.+\ket|+-+-+>+\ket|+--++>\)
\label{eq:app_gs_onehalf}
\end{align}

\subsection{Tensor network representation}
\label{app:mps}

In this section we will write the GS of the site-star Heisenberg model
as a matrix product state (MPS). Given the Matrioshka structure of the
described RG procedure, let us start writing the final state in the
$Z$ component basis
$\ket|\bfS_\ell,\mathfrak{S}_\ell>\equiv\ket|\mathfrak{S}_\ell>$ using
closure relations:
\begin{equation}
  \ket|\mathfrak{S}_\ell> 
  = \sum_{\mathfrak{S}_{\ell-1},\,\mathcal{S}_\ell}
  \braket{\mathfrak{S}_{\ell-1};\,\mathcal{S}_\ell}{\mathfrak{S}_\ell}\,
  \ket|\mathfrak{S}_{\ell-1};\,\mathcal{S}_\ell> 
  \equiv \sum_{\mathfrak{S}_{\ell-1},\,\mathcal{S}_\ell}
  C^{\mathfrak{S}_\ell}_{\,\mathfrak{S}_{\ell-1}\,\mathcal{S}_\ell}\,
  \ket|\mathfrak{S}_{\ell-1};\,\mathcal{S}_\ell>\,.
\end{equation}
We can iterate this procedure $\ell-1$ times:
\begin{equation}
  \ket|\mathfrak{S}_\ell> = \sum_{\substack{\mathfrak{S}_1 \cdots
    \mathfrak{S}_{\ell-1}\\\mathcal{S}_1\cdots \mathcal{S}_\ell}}
C^{\mathfrak{S}_{1}}_{\mathfrak{S}_{0}\,\mathcal{S}_{1}}
C^{\mathfrak{S}_{2}}_{\mathfrak{S}_{1}\,\mathcal{S}_{2}}
\cdots 
C^{\mathfrak{S}_{\ell}}_{\mathfrak{S}_{\ell-1}\,\mathcal{S}_{\ell}}
\ket|s_{0};\,\mathcal{S}_1;\,\dots;\,\mathcal{S}_\ell>\,.
\label{eq:app_protomps}
\end{equation}
where we have rewritten $\mathfrak{S}_0\equiv s_0$ as the central
physical spin. Observe that the interchange of the indices

$$C^{M_{\mathfrak{S}_{m}}}_{M_{\mathfrak{S}_{m-1}}M_{\mathcal{S}_{m}}}\longrightarrow
\Gamma^{M_{\mathcal{S}_{m}}}_{M_{\mathfrak{S}_{m-1}}M_{\mathfrak{S}_{m}}}, $$
allows to see Eq.\eqref{eq:app_protomps} as a product of matrices:
\begin{align}
  \ket|\mathfrak{S}_\ell>&
  =\sum_{s_0,\mathcal{S}_1,\dots\mathcal{S}_\ell}
  \sum_{\mathfrak{S}_1,\dots\mathfrak{S}_{\ell-1}}
  \Gamma^{\mathcal{S}_1}_{s_0\mathfrak{S}_1}
  \Gamma^{\mathcal{S}_2}_{\mathfrak{S}_1\mathfrak{S}_2} \dots
  \Gamma^{\mathcal{S}_\ell}_{\mathfrak{S}_{\ell-1}\mathfrak{S}_\ell}
  \ket|s_0,\mathcal{S}_1,\mathcal{S}_2,\dots,\mathcal{S}_\ell>\\
  &=\sum_{s_0,\mathcal{S}_1,\dots\mathcal{S}_\ell}
  \bm{\Gamma}^{s_0}\bm{\Gamma}^{\mathcal{S}_1}\bm{\Gamma}^{\mathcal{S}_2}\hdots
  \bm{\Gamma}^{\mathcal{S}_\ell}
  \ket|s_0,\mathcal{S}_1,\mathcal{S}_2,\dots,\mathcal{S}_\ell>,
\label{eq:app_MPS}
\end{align}
where we have added the boundary vector
$\Gamma^{\mathcal{S}_1}_{s_0\mathfrak{S}_1}$. Expression
\eqref{eq:app_MPS} matrix-product state (MPS) with open boundary
conditions consisting of $\ell+1$ spins, with the initial spin
representing the physical spin at site 0 and the subsequent spins
being obtained as maximal composition of the spins lying at the same
level $m=1,\dots,\ell$. The bond dimension is given by the
renormalized spins and therefore alternates between $\mathfrak{S}=1/2$
and $\mathfrak{S}=(n_B-1)/2$. As a consequence, the matrices
$\bm{\Gamma}$ consist of appropriate Clebsch-Gordan coefficients and
are rectangular $2\times n_B$.

To gain insight, let us consider the system of $n_B=2$ branches. In
this case, the physical spins are $\mathcal{S}=1$ and the bond
dimension is always $2$, since $\mathfrak{S}=1/2$. As can be seen in
Fig. \ref{fig:cleb2}, we obtain the MPS tensors by reading the
suitable Clebsch-Gordan coefficients:

\begin{figure}[ht]
  \centering
\setlength{\unitlength}{0.9mm}
  \begin{picture}(64,32)(100,76)           
    \put(116,100){\line(-1,0){16}}
    \put(116,100){\line(0,1){8}}
    \put(100,96){\line(1,0){24}}
    \put(100,96){\line(0,1){4}}
    \put(124,108){\line(-1,0){8}}
    \put(124,108){\line(0,-1){12}}
    \multiput(116,100)(2,0){4}{\line(1,0){1}}
    \multiput(116,100)(0,-2){2}{\line(0,-1){1}}
    \put(108,88){\line(1,0){32}}
    \put(108,88){\line(0,1){8}}
    \put(140,104){\line(-1,0){16}}
    \put(140,104){\line(0,-1){16}}
    \multiput(124,96)(2,0){8}{\line(1,0){1}}
    \multiput(124,96)(0,-2){4}{\line(0,-1){1}}
    \put(124,80){\line(1,0){32}}
    \put(124,80){\line(0,1){8}}
    \put(156,96){\line(-1,0){16}}
    \put(156,96){\line(0,-1){16}}
    \multiput(140,88)(2,0){8}{\line(1,0){1}}
    \multiput(140,88)(0,-2){4}{\line(0,-1){1}}
    \put(140,76){\line(1,0){24}}
    \put(140,76){\line(0,1){4}}
    \put(164,88){\line(-1,0){8}}
    \put(164,88){\line(0,-1){12}}
    \multiput(156,80)(2,0){4}{\line(1,0){1}}
    \multiput(156,80)(0,-2){2}{\line(0,-1){1}}
    \put(108,104){\makebox(0,0){\normalsize 1$\, \times \,$1/2}}
    \put(120,106){\makebox(0,0){3/2}}
    \put(120,102){\makebox(0,0){3/2}}
    \put(128,102){\makebox(0,0){3/2}}
    \put(136,102){\makebox(0,0){1/2}}
    \put(104,98){\makebox(0,0){1}}
    \put(112,98){\makebox(0,0){1/2}}
    \put(120,98){\makebox(0,0){1}}
    \put(128,98){\makebox(0,0){1/2}}
    \put(136,98){\makebox(0,0){1/2}}
    \put(112,94){\makebox(0,0){1}}
    \put(120,94){\makebox(0,0){-1/2}}
    \put(128,94){\makebox(0,0){1/3}}
    \put(136,94){\makebox(0,0){\textcolor{blue}{2/3}}}
    \put(144,94){\makebox(0,0){3/2}}
    \put(152,94){\makebox(0,0){1/2}}
    \put(112,90){\makebox(0,0){0}}
    \put(120,90){\makebox(0,0){1/2}}
    \put(128,90){\makebox(0,0){2/3}}
    \put(136,90){\makebox(0,0){\textcolor{blue}{-1/3}}}
    \put(144,90){\makebox(0,0){-1/2}}
    \put(152,90){\makebox(0,0){-1/2}}
    \put(128,86){\makebox(0,0){0}}
    \put(136,86){\makebox(0,0){-1/2}}
    \put(144,86){\makebox(0,0){2/3}}
    \put(152,86){\makebox(0,0){\textcolor{blue}{1/3}}}
    \put(160,86){\makebox(0,0){3/2}}
    \put(128,82){\makebox(0,0){-1}}
    \put(136,82){\makebox(0,0){1/2}}
    \put(144,82){\makebox(0,0){1/3}}
    \put(152,82){\makebox(0,0){\textcolor{blue}{-2/3}}}
    \put(160,82){\makebox(0,0){-3/2}}
    \put(144,78){\makebox(0,0){-1}}
    \put(152,78){\makebox(0,0){-1/2}}
    \put(160,78){\makebox(0,0){1}}
  \end{picture}
 
  \caption{Clebsch-Gordan coefficients $1\otimes1/2$}
  \label{fig:cleb2}
\end{figure}
\begin{eqnarray}
\bm{\Gamma}^{1} = \begin{pmatrix}
     0 & \sqrt{\frac{2}{3}}\\
     0 & 0
\end{pmatrix},\quad 
\bm{\Gamma}^{0} = \begin{pmatrix} -\sqrt{\frac{1}{3}} & 0\\ 0 &
  \sqrt{\frac{1}{3}}
\end{pmatrix},\quad        
\bm{\Gamma}^{-1} = \begin{pmatrix}
    0 & 0 \\
    -\sqrt{\frac{2}{3}} & 0 
\end{pmatrix},\quad 
\end{eqnarray}
That correspond to the well known tensors of the AKLT state. Observe,
however the crucial difference: whereas in this case the physical and
virtual indices of the MPS description aligns with the physical
degrees of freedom of the AKLT Hamiltonian- which describes the
dynamics of spin $1$ particles, our case is the opposite. The physical
indices of the MPS do not appear in the spin $1/2$ inhomogeneous
Hamiltonian described in the main text. They are indeed virtual spins
formed to minimise the energy.

Now we may consider the star of $n_B=3$ branches. In this case, the
MPS description involves spins $\mathcal{S}=3/2$, and the matrices are
rectangular, since we have the alternation $\mathfrak{S}=1/2$ and
$\mathfrak{S}=1$. Hence, we extract the corresponding tensors of the odd spins from the Clebchs-Gordan coeficients $1/2\times 3/2$ (see Fig. \ref{fig:cleb3} (a)):
\begin{equation}
\bm{\Gamma}^{-3/2} =\left(
\begin{array}{ccc}
 0 & 0 & 0 \\
 -\frac{\sqrt{3}}{2} & 0 & 0 \\
\end{array}
\right),\quad \bm{\Gamma}^{-1/2} =\left(
\begin{array}{ccc}
 \frac{1}{2} & 0 & 0 \\
 0 & -\frac{1}{\sqrt{2}} & 0 \\
\end{array}
\right),\quad \bm{\Gamma}^{1/2} =\left(
\begin{array}{ccc}
 0 & \frac{1}{\sqrt{2}} & 0 \\
 0 & 0 & -\frac{1}{2} \\
\end{array}
\right),\quad \bm{\Gamma}^{3/2} =\left(
\begin{array}{ccc}
 0 & 0 & \frac{\sqrt{3}}{2} \\
 0 & 0 & 0 \\
\end{array}
\right).
\end{equation}

On the other hand, the tensors describing spins at even positions are obtainbed from the Clebsch-Gordan coefficients $1\times 3/2$ (see Fig. \ref{fig:cleb3a}), taking the form:
\begin{equation}
\tilde{\bm{\Gamma}}^{-3/2} =\left(
\begin{array}{cc}
 0 & 0 \\
 0 & 0 \\
 \frac{1}{\sqrt{2}} & 0 \\
\end{array}
\right),\quad \tilde{\bm{\Gamma}}^{-1/2} =\left(
\begin{array}{cc}
 0 & 0 \\
 -\frac{1}{\sqrt{3}} & 0 \\
 0 & \frac{1}{\sqrt{6}} \\
\end{array}
\right),\quad \tilde{\bm{\Gamma}}^{1/2} =\left(
\begin{array}{cc}
 \frac{1}{\sqrt{6}} & 0 \\
 0 & -\frac{1}{\sqrt{3}} \\
 0 & 0 \\
\end{array}
\right),\quad \tilde{\bm{\Gamma}}^{3/2} =\left(
\begin{array}{cc}
 0 & \frac{1}{\sqrt{2}} \\
 0 & 0 \\
 0 & 0 \\
\end{array}
\right)
\end{equation}

Crucially,  notice that the product of two consecutive tensors
$\bm{\Gamma}^{\mathcal{S}_1}\tilde{\bm{\Gamma}}^{\mathcal{S}_2}$
yields non-zero $2\times 2$ matrices provided
$\mathcal{S}_1+\mathcal{S}_2=\pm1, 0$. This suggests that a pair of
neighboring spins behave as a spin $1$. To further progress, we
compute the transfer matrix of the product
\begin{equation} T\equiv\sum_{\mathcal{S}_1\mathcal{S}_2}
  \(\bm{\Gamma}^{\mathcal{S}_1}\otimes(\bm{\Gamma}^{\mathcal{S}_1})^*\)
  \(\tilde{\bm{\Gamma}}^{\mathcal{S}_2}\otimes(\tilde{\bm{\Gamma}}^{\mathcal{S}_2})^*\)=
  \(\begin{array}{cccc}
 \frac{7}{12} & 0 & 0 & \frac{5}{12} \\
 0 & \frac{1}{6} & 0 & 0 \\
 0 & 0 & \frac{1}{6} & 0 \\
 \frac{5}{12} & 0 & 0 & \frac{7}{12} \\
\end{array}\).
\end{equation}

Proceeding like before, we read the coefficients from the
Clebsch-Gordan table (see Fig.\ref{fig:cleb3}).

\begin{figure}[ht]
\setlength{\unitlength}{0.9mm}
\hspace{-3cm}\subfigure{
    \begin{minipage}{0.45\textwidth}
  \begin{picture}(80,40)(84,110)
\put(84,138){\line(1,0){24}}
\put(84,138){\line(0,1){4}}
\put(100,142){\line(-1,0){16}}
\put(100,142){\line(0,1){8}}
\put(108,150){\line(-1,0){8}}
\put(108,150){\line(0,-1){12}}
\multiput(100,142)(2,0){4}{\line(1,0){1}}
\multiput(100,142)(0,-2){2}{\line(0,-1){1}}
\put(92,130){\line(1,0){32}}
\put(92,130){\line(0,1){8}}
\put(124,146){\line(-1,0){16}}
\put(124,146){\line(0,-1){16}}
\multiput(108,138)(2,0){8}{\line(1,0){1}}
\multiput(108,138)(0,-2){4}{\line(0,-1){1}}
\put(108,122){\line(1,0){32}}
\put(108,122){\line(0,1){8}}
\put(140,138){\line(-1,0){16}}
\put(140,138){\line(0,-1){16}}
\multiput(124,130)(2,0){8}{\line(1,0){1}}
\multiput(124,130)(0,-2){4}{\line(0,-1){1}}
\put(124,114){\line(1,0){32}}
\put(124,114){\line(0,1){8}}
\put(156,130){\line(-1,0){16}}
\put(156,130){\line(0,-1){16}}
\multiput(140,122)(2,0){8}{\line(1,0){1}}
\multiput(140,122)(0,-2){4}{\line(0,-1){1}}
\put(140,110){\line(1,0){24}}
\put(140,110){\line(0,1){4}}
\put(164,122){\line(-1,0){8}}
\put(164,122){\line(0,-1){12}}
\multiput(156,114)(2,0){4}{\line(1,0){1}}
\multiput(156,114)(0,-2){2}{\line(0,-1){1}}
\put(90,146){\makebox(0,0){\normalsize 3/2$\, \times \,$1/2}}
\put(104,148){\makebox(0,0){2}}
\put(104,144){\makebox(0,0){2}}
\put(112,144){\makebox(0,0){2}}
\put(120,144){\makebox(0,0){1}}
\put(88,140){\makebox(0,0){3/2}}
\put(96,140){\makebox(0,0){1/2}}
\put(104,140){\makebox(0,0){1}}
\put(112,140){\makebox(0,0){1}}
\put(120,140){\makebox(0,0){1}}
\put(96,136){\makebox(0,0){3/2}}
\put(104,136){\makebox(0,0){-1/2}}
\put(112,136){\makebox(0,0){1/4}}
\put(120,136){\makebox(0,0){\textcolor{blue}{3/4}}}
\put(128,136){\makebox(0,0){2}}
\put(136,136){\makebox(0,0){1}}
\put(96,132){\makebox(0,0){1/2}}
\put(104,132){\makebox(0,0){1/2}}
\put(112,132){\makebox(0,0){3/4}}
\put(120,132){\makebox(0,0){\textcolor{blue}{-1/4}}}
\put(128,132){\makebox(0,0){0}}
\put(136,132){\makebox(0,0){0}}
\put(112,128){\makebox(0,0){1/2}}
\put(120,128){\makebox(0,0){-1/2}}
\put(128,128){\makebox(0,0){1/2}}
\put(136,128){\makebox(0,0){\textcolor{blue}{1/2}}}
\put(144,128){\makebox(0,0){2}}
\put(152,128){\makebox(0,0){1}}
\put(112,124){\makebox(0,0){-1/2}}
\put(120,124){\makebox(0,0){1/2}}
\put(128,124){\makebox(0,0){1/2}}
\put(136,124){\makebox(0,0){\textcolor{blue}{-1/2}}}
\put(144,124){\makebox(0,0){-1}}
\put(152,124){\makebox(0,0){-1}}
\put(128,120){\makebox(0,0){-1/2}}
\put(136,120){\makebox(0,0){-1/2}}
\put(144,120){\makebox(0,0){3/4}}
\put(152,120){\makebox(0,0){\textcolor{blue}{1/4}}}
\put(160,120){\makebox(0,0){2}}
\put(128,116){\makebox(0,0){-3/2}}
\put(136,116){\makebox(0,0){1/2}}
\put(144,116){\makebox(0,0){1/4}}
\put(152,116){\makebox(0,0){\textcolor{blue}{-3/4}}}
\put(160,116){\makebox(0,0){-2}}
\put(144,112){\makebox(0,0){-3/2}}
\put(152,112){\makebox(0,0){-1/2}}
\put(160,112){\makebox(0,0){1}}
\end{picture}\end{minipage}}
\hspace{-3cm}\subfigure{
    \begin{minipage}{0.45\textwidth}
      \begin{picture}(120,55)(44,41)
    \put(44,84){\line(1,0){24}}
\put(44,84){\line(0,1){4}}
\put(60,88){\line(-1,0){16}}
\put(60,88){\line(0,1){8}}
\put(68,96){\line(-1,0){8}}
\put(68,96){\line(0,-1){12}}
\multiput(60,88)(2,0){4}{\line(1,0){1}}
\multiput(60,88)(0,-2){2}{\line(0,-1){1}}
\put(52,76){\line(1,0){32}}
\put(52,76){\line(0,1){8}}
\put(84,92){\line(-1,0){16}}
\put(84,92){\line(0,-1){16}}
\multiput(68,84)(2,0){8}{\line(1,0){1}}
\multiput(68,84)(0,-2){4}{\line(0,-1){1}}
\put(68,64){\line(1,0){44}}
\put(68,64){\line(0,1){12}}
\put(112,84){\line(-1,0){28}}
\put(112,84){\line(0,-1){20}}
\multiput(84,76)(2,0){14}{\line(1,0){1}}
\multiput(84,76)(0,-2){6}{\line(0,-1){1}}
\put(94,52){\line(1,0){46}}
\put(94,52){\line(0,1){12}}
\put(140,72){\line(-1,0){28}}
\put(140,72){\line(0,-1){20}}
\multiput(112,64)(2,0){14}{\line(1,0){1}}
\multiput(112,64)(0,-2){6}{\line(0,-1){1}}
\put(122,44){\line(1,0){34}}
\put(122,44){\line(0,1){8}}
\put(156,60){\line(-1,0){16}}
\put(156,60){\line(0,-1){16}}
\multiput(140,52)(2,0){8}{\line(1,0){1}}
\multiput(140,52)(0,-2){4}{\line(0,-1){1}}
\put(140,40){\line(1,0){24}}
\put(140,40){\line(0,1){4}}
\put(164,52){\line(-1,0){8}}
\put(164,52){\line(0,-1){12}}
\multiput(156,44)(2,0){4}{\line(1,0){1}}
\multiput(156,44)(0,-2){2}{\line(0,-1){1}}
\put(52,92){\makebox(0,0){\normalsize 3/2$\, \times \,$1}}
\put(64,94){\makebox(0,0){5/2}}
\put(64,90){\makebox(0,0){5/2}}
\put(72,90){\makebox(0,0){5/2}}
\put(80,90){\makebox(0,0){3/2}}
\put(48,86){\makebox(0,0){3/2}}
\put(56,86){\makebox(0,0){1}}
\put(64,86){\makebox(0,0){1}}
\put(72,86){\makebox(0,0){3/2}}
\put(80,86){\makebox(0,0){3/2}}
\put(56,82){\makebox(0,0){3/2}}
\put(64,82){\makebox(0,0){0}}
\put(72,82){\makebox(0,0){2/5}}
\put(80,82){\makebox(0,0){3/5}}
\put(89,82){\makebox(0,0){5/2}}
\put(99,82){\makebox(0,0){3/2}}
\put(108,82){\makebox(0,0){1/2}}
\put(56,78){\makebox(0,0){1/2}}
\put(64,78){\makebox(0,0){1}}
\put(72,78){\makebox(0,0){3/5}}
\put(80,78){\makebox(0,0){-2/5}}
\put(89,78){\makebox(0,0){1/2}}
\put(99,78){\makebox(0,0){1/2}}
\put(108,78){\makebox(0,0){1/2}}
\put(72,74){\makebox(0,0){3/2}}
\put(80,74){\makebox(0,0){-1}}
\put(89,74){\makebox(0,0){1/10}}
\put(99,74){\makebox(0,0){2/5}}
\put(108,74){\makebox(0,0){\textcolor{blue}{1/2}}}
\put(72,70){\makebox(0,0){1/2}}
\put(80,70){\makebox(0,0){0}}
\put(89,70){\makebox(0,0){3/5}}
\put(99,70){\makebox(0,0){1/15}}
\put(108,70){\makebox(0,0){\textcolor{blue}{-1/3}}}
\put(117,70){\makebox(0,0){5/2}}
\put(127,70){\makebox(0,0){3/2}}
\put(136,70){\makebox(0,0){1/2}}
\put(72,66){\makebox(0,0){-1/2}}
\put(80,66){\makebox(0,0){1}}
\put(89,66){\makebox(0,0){3/10}}
\put(99,66){\makebox(0,0){-8/15}}
\put(108,66){\makebox(0,0){\textcolor{blue}{1/6}}}
\put(117,66){\makebox(0,0){-1/2}}
\put(127,66){\makebox(0,0){-1/2}}
\put(136,66){\makebox(0,0){-1/2}}
\put(99,62){\makebox(0,0){1/2}}
\put(108,62){\makebox(0,0){-1}}
\put(117,62){\makebox(0,0){3/10}}
\put(127,62){\makebox(0,0){8/15}}
\put(136,62){\makebox(0,0){\textcolor{blue}{1/6}}}
\put(99,58){\makebox(0,0){-1/2}}
\put(108,58){\makebox(0,0){0}}
\put(117,58){\makebox(0,0){3/5}}
\put(127,58){\makebox(0,0){-1/15}}
\put(136,58){\makebox(0,0){\textcolor{blue}{-1/3}}}
\put(144,58){\makebox(0,0){5/2}}
\put(152,58){\makebox(0,0){3/2}}
\put(99,54){\makebox(0,0){-3/2}}
\put(108,54){\makebox(0,0){1}}
\put(117,54){\makebox(0,0){1/10}}
\put(127,54){\makebox(0,0){-2/5}}
\put(136,54){\makebox(0,0){\textcolor{blue}{1/2}}}
\put(144,54){\makebox(0,0){-3/2}}
\put(152,54){\makebox(0,0){-3/2}}
\put(127,50){\makebox(0,0){-1/2}}
\put(136,50){\makebox(0,0){-1}}
\put(144,50){\makebox(0,0){3/5}}
\put(152,50){\makebox(0,0){2/5}}
\put(160,50){\makebox(0,0){5/2}}
\put(127,46){\makebox(0,0){-3/2}}
\put(136,46){\makebox(0,0){0}}
\put(144,46){\makebox(0,0){2/5}}
\put(152,46){\makebox(0,0){-3/5}}
\put(160,46){\makebox(0,0){-5/2}}
\put(144,42){\makebox(0,0){-3/2}}
\put(152,42){\makebox(0,0){-1}}
\put(160,42){\makebox(0,0){1}}
\vspace{1cm}
  \end{picture}\end{minipage}
  }
\caption{Clebsch-Gordan coefficients. Left: $3/2\times 1/2$ used for obtaining $\mathbf{\Gamma}$ tensors. Right: $3/2 \times 1$ used for obtaining $\tilde{\mathbf{\Gamma}}$ tensors.}
   \label{fig:cleb3}
\end{figure}


\begin{thebibliography}{99}
\bibitem{note1} Observe that the
Jordan-Wigner transformation does not map this model to an XX spin
system.
\end{thebibliography}
\end{document}